\documentclass[a4paper,11pt]{article}

\usepackage{jheppub} 

\usepackage[T1]{fontenc} 
\usepackage{amsmath}
\usepackage{amssymb}
\usepackage{graphicx}
\usepackage{hyperref}
\usepackage{xcolor}
\usepackage{mathtools}
\usepackage{cancel}
\usepackage[normalem]{ulem}
\usepackage{changes}

\newcommand{\E}{\text{E}}
\renewcommand{\P}{\text{P}}
\newcommand{\df}{\text{d}}
\newcommand{\p}{\partial}

\newcommand{\ov}{\over}
\newcommand{\al}[1]{\begin{align}#1\end{align}}

\newcommand{\pn}[1]{\left(#1\right)}
\newcommand{\Pn}[1]{\bigl(#1\bigr)}
\newcommand{\br}[1]{\left\{#1\right\}}

\newcommand{\sqbr}[1]{\left[#1\right]}
\newcommand{\Sqbr}[1]{\bigl[#1\bigr]}
\newcommand{\fn}[1]{\!\left(#1\right)}
\newcommand{\Fn}[1]{\!\bigl(#1\bigr)}
\newcommand{\ab}[1]{\left|#1\right|}

\newcommand{\nn}{\nonumber\\}
\newcommand{\tx}{\text}

\newcommand{\oc}[1]{\overset{\circ}{#1}{}}	

\newcommand{\wt}{\widetilde}

\begin{document} 

\title{\boldmath Ultraviolet Sensitivity in Higgs-Starobinsky Inflation}

\preprint{CERN-TH-2023-040}


\author[a,b]{Sung Mook Lee,} 
\author[c]{Tanmoy Modak,}
\author[d]{Kin-ya Oda,}
\author[e]{and Tomo Takahashi}


\affiliation[a]{Department of Physics \& IPAP \& Lab for Dark Universe, Yonsei University, Seoul 03722, Korea}
\affiliation[b]{Theoretical Physics Department, CERN, 1211 Geneva 23, Switzerland
}
\affiliation[c]{Institut f{\"u}r Theoretische Physik, Universit{\"a}t Heidelberg, 69120 Heidelberg, Germany}
\affiliation[d]{Department of Mathematics, Tokyo Woman's Christian University, Tokyo 167-8585, Japan}
\affiliation[e]{Department of Physics, Saga University, Saga 840-8502, Japan}

\emailAdd{sungmook.lee@yonsei.ac.kr}
\emailAdd{tanmoyy@thphys.uni-heidelberg.de}
\emailAdd{odakin@lab.twcu.ac.jp}
\emailAdd{tomot@cc.saga-u.ac.jp}

\abstract{	
The general scalar-tensor theory that includes all the dimension-four terms has parameter regions that can produce successful inflation consistent with cosmological observations. This theory is in fact the same as the Higgs-Starobinsky inflation, when the scalar is identified with the Standard Model Higgs boson.
We consider possible dimension-six operators constructed from non-derivative terms of the scalar field and the Ricci scalar as perturbations. We investigate how much suppression is required for these operators to avoid disrupting the successful inflationary predictions.
To ensure viable cosmological predictions, the suppression scale for the sixth power of the scalar should be as high as the Planck scale. For the other terms, much smaller scales are sufficient.
}

\maketitle
\flushbottom

\section{Introduction}

Inflation is now an essential part of standard cosmology, but we still do not know what caused the quasi-de Sitter phase in the early universe. To realize this inflationary phase, we require at least one scalar degree of freedom, the `inflaton,' in the beyond Standard Model (BSM) sector.
One exception is Higgs inflation~\cite{Salopek:1988qh,Bezrukov:2007ep,Lee:2020yaj,Cheong:2021vdb}, where the Standard Model (SM) Higgs boson plays the role of the inflaton. However, this ``vanilla'' Higgs inflation model suffers from a unitarity problem, which still necessitates a new degree of freedom to recover perturbativity up to the Planck scale~\cite{Giudice:2010ka}. More precisely, the inflation itself is not spoiled~\cite{Bezrukov:2010jz}, but its reheating process involves higher momenta than the cutoff scale~\cite{Ema:2016dny}.

On the other hand, Starobinsky inflation~\cite{Starobinsky:1980te}, or $R^2$-inflation, is the first and one of the most minimal models of inflation. It still fits within the sweet spot of current measurements on the spectral index $n_s$ and the tensor-to-scalar ratio $r$ at around 60 $e$-folds before the end of inflation. From the perspective of modified general relativity (GR), $R^2$ inflation can be regarded as a special case of $f(R)$-theories (for a review, see e.g., \cite{Sotiriou:2008rp,DeFelice:2010aj}), which generally introduce a new scalar degree of freedom, named the \textit{scalaron}, in the metric formalism.\footnote{
	In this paper, we assume the metric formalism and not the Palatini one.
	In the latter, the $R^2$ term does not introduce any new dynamical degree of freedom and we still need to add another scalar degree of freedom to realize the inflation~\cite{Sotiriou:2008rp,Antoniadis:2018ywb,Cheong:2021kyc}.
}

Considering general BSM theories in particle physics at high energies, it is plausible to have other scalars in the spectrum than the scalaron degree of freedom; see e.g.\ Ref.~\cite{Cicoli:2011zz}. Because we already have at least one scalar in the Standard Model, the Higgs, we must include at least one (possibly non-minimally coupled) scalar $\chi$ to write down the most general action of an $ R^{2} $-type inflation. Throughout this paper, $ \chi $ stands for an arbitrary scalar field, albeit we call it the Higgs for concreteness.

One explicit example of such a scalar extension of the $R^2$ inflation is the Higgs-$ R^{2}$ inflation~\cite{Salvio:2015kka,Salvio:2016vxi,Ema:2017rqn,Gorbunov:2018llf,He:2018gyf,He:2018mgb,Gundhi:2018wyz,Ema:2019fdd}, which has been extended to take into account all the possible terms up to dimension four in the general $ f( R , \chi) $ type of theory~\cite{Canko:2019mud}. (In fact, the role of the mass term for the scalar is usually overlooked in the literature. We briefly discuss these aspects in Appendix~\ref{App:mass}.)
One can also regard the Higgs-$R^2$ model as an extension of the vanilla Higgs inflation to cure the above-mentioned unitarity problem during the reheating.

In general, large-field inflationary models suffer from a naturalness problem. A well-known example is the $ \eta $-problem, which questions how the inflaton remains massless during inflation without being protected by any symmetry; see e.g.\ Ref.~\cite{Baumann:2014nda}. In other words, we need to explain what assures the approximate shift symmetry that forbids the mass term in the inflaton potential.\footnote{
The absence of the renormalized mass term is sometimes called the classical scale invariance or the classical conformality. Recently it has been shown that such an absence can be understood as a realization of the multicritical point principle~\cite{Kawai:2021lam}.
}
$ R^{2} $, Higgs, and Higgs-$ R^{2} $ inflation models all implicitly assume this approximate shift symmetry at the super-Planckian regime $\chi\gtrsim M_\P$ such that the energy density is much smaller than the Planck scale: $ V(\chi) \ll M_\P^{4} $. This smallness of the energy density is sometimes used to justify the larger field values than the cutoff scale in the effective theory of inflation~\cite{Linde:1983gd}.\footnote{
Note however that e.g.\ the so-called distance conjecture says: ``We cannot have a slow roll inflation where the distance in the scalar moduli space is much bigger than Planck length and still use the same effective field theory''~\cite{Ooguri:2006in}.
}
To realize this approximate shift symmetry at the super-Planckian field values, careful consideration of the higher dimensional terms in the potential plays a key role.

Theoretically, it is favored that an inflation model has ultraviolet (UV) insensitivity. For example, it was shown that the cosmological observables of single-field inflation with large non-minimal coupling are insensitive to the addition of dimension-6 operators in metric formalism~\cite{Jinno:2019und} barring the unitarity problem during the reheating. In this work, we ask to what extent inflationary dynamics and observational predictions depend on the presence of a dimension-six operator and to what extent such operators should be suppressed in order not to spoil the consistency with observations.

In this direction, previous literature mainly concerns either $ R^{n} $ (especially $ n=3 $) type of correction to the $ R^{2} $-inflation~\cite{Huang:2013hsb,Cheong:2020rao,Ivanov:2021chn}, or $ \chi^{6} $ term or $ \chi^{4} R $ term in the Higgs inflation~\cite{Jinno:2019und}. In this work, we generalize these arguments by considering all possible dimension-six no-extra-derivative operators constructed from $\chi$ and $R$ such as $ \chi^{2} R^{2} $, $ \chi^{4}R $, and $ \chi^{6} $ in the Higgs-$ R^{2} $ inflation.\footnote{
For simplicity, we assume the $Z_2$ symmetry $\chi\to-\chi$, with the gauge invariance of the Higgs in mind.
}
\footnote{
	There may be other kinds of dimension-6 operators which involves derivatives like $ R (\partial \chi)^{2}$. The effects of derivative couplings are generally expected to be more suppressed than non-derivative ones in the slow-roll inflation. Detailed consideration requires another machinery beyond the scope of this work and we will only consider polynomial terms of $\chi$ and $R$ within $f(R,\chi)$ framework.
}

This work is organized as follows. In Section~\ref{Sec2}, we review the general $ f(R, \chi) $ theory and set our notation. We also summarize our analysis method to investigate the prediction of inflation models in the framework of the $ f(R, \chi) $ theory. In Section~\ref{Sec3}, we will consider the effects of higher dimensional operators on the scalar potential and cosmological observables. Then we conclude in Section~\ref{Sec:Conclusion and Discussion}.

\section{$ f(R, \chi) $ theory and Higgs-$ R^{2} $ inflation} \label{Sec2}

In this section, we review inflation models in the general $ f (R, \chi) $ theory. First, we will briefly discuss the general $ f\fn{R,\chi} $ theory and its dual scalar-tensor-theory picture containing another scalar field, the scalaron, other than the Higgs, with canonical Einstein gravity. Second, considering all possible operators up to dimension four automatically gives Higgs-$ R^{2} $ theory of inflation. Albeit we consider a real singlet scalar $ \chi $ in this paper, generalization to SM Higgs or other charged scalar should be straightforward.

\subsection{General Potential in $ f\fn{R,\chi} $ theory}

We shortly review transformations of $ f\fn{R,\chi} $ theory in the metric formalism where the affine connection is assumed to be the Levi-Civita connection {\it a priori}; see e.g.\ Ref.~\cite{Sotiriou:2008rp} for a more comprehensive review.

We first consider a general  $ f\fn{R,\chi} $ gravity with the action
\begin{align}
	S  = \int\df^{4}x \sqrt{-g} \left[ \frac{M_\P^{2}}{2}  f\fn{R,\chi} + \mathcal{L}_{\text{mat}}\fn{\chi}	\right],
		\label{starting action}
\end{align}
where $M_\P=1/\sqrt{8\pi G} \simeq 2.4\times10^{18}\,\text{GeV}$ is the reduced Planck mass.
We assume that there is a frame in which the matter sector is canonically normalized,
\begin{align}
	\mathcal{L}_{\text{mat}} = -\frac{1}{2} \pn{\p\chi}^2 - U\fn{\chi},
		\label{canonical scalar Lagrangian}
\end{align}
where $ U(\chi) $ is an arbitrary potential of the $ \chi $ field (without containing its derivatives).

In pure GR with a `canonically-coupled' scalar, $f\fn{R,\chi}=R$, it is known that the conformal mode of the metric has a wrong-sign kinetic term; see Appendix~\ref{conformal mode} for more detail. In such a model, even if one Wick-rotates the time coordinate, either the conformal mode or the matter field exponentially grows in both the imaginary-time directions, making Euclideanization impossible; see e.g.\ Refs.~\cite{Gibbons:1978ac,tHooft:2011aa}. In any case, no symmetry forbids non-minimal couplings among $R$ and $\chi$, and hence, even if we drop them at the tree level by hand, they are generated via loop corrections. Therefore we do not assume the pure GR with the canonically coupled scalar.
In practice, this means that we do not drop the $R^2$ and $\chi^2R$ terms by hand at the dimension-four truncation. Especially, the following assumption becomes relevant for the following discussion:
\al{
{\p^2f\fn{R,\chi}\ov\p R^2}\neq0.
	\label{second derivative of f}
}

Introducing an auxiliary field $ \Psi $, we may write as
\begin{align}
	S = \int\df^{4}x \sqrt{-g} \left[ \frac{M_\P^{2}}{2} \pn{ f\fn{\Psi,\chi} + \frac{\p f\fn{\Psi,\chi}}{\p\Psi} \left(R -\Psi	\right) } +  \mathcal{L}_{\text{mat}}\fn{\chi}  \right].
\end{align}
A variation with respect to $ \Psi $ gives the constraint $ \Psi = R $ as long as Eq.~\eqref{second derivative of f} holds.
We trade $\Psi$ for a physical degree of freedom:
\begin{align}
	\Phi = \frac{\p f\fn{\Psi,\chi}}{\p\Psi}.
\end{align}
Now the action in the Jordan frame reads
\begin{align}
	S = \int d^{4}x \sqrt{-g} \left[ \frac{M_\P^{2}}{2}\Phi R - V\fn{\Phi,\chi} + \mathcal{L}_{\text{mat}}\fn{\chi} \right],
\end{align}
where
\begin{align}
	V(\Phi,\chi) := \frac{M_\P^{2}}{2} \Sqbr{ \Psi(\Phi) \Phi - f\Fn{\Psi (\Phi),\chi} }.
	\label{V given}
\end{align}

To obtain the action in the Einstein frame, we redefine the metric field by
\begin{align}
g_{\E\mu\nu}
	&= \Phi g_{\mu\nu}
\end{align}
so that
\begin{align}
	S = \int d^{4}x \sqrt{-g_\E} \left[  \frac{M_\P^{2}}{2} R_\E  - \frac{3 M_\P^{2}}{4 \Phi^{2}} \pn{\partial \Phi}_\E^2
	- \frac{1}{2 \Phi}\pn{\p\chi}_\E^2 - \frac{1}{\Phi^{2}} \Pn{V\fn{\Phi,\chi} + U\fn{\chi} } \right],
\end{align}
where $\pn{\p\varphi}^2_\E:=g_\E^{\mu\nu}\p_\mu\varphi\p_\nu\varphi$ etc.; see Appendix~\ref{conformal mode}.
Note that the vanilla Higgs inflation $f\fn{R,\chi}=\fn{1+{\chi^2\ov M_\P^2}}R$ does not satisfy the condition~\eqref{second derivative of f}, and hence $\Phi=1+{\chi^2\ov M_\P^2}$ does not contain the extra scalaron degree of freedom other than $\chi$.

Canonicalizing the inflaton field $ \Phi $ with
\begin{align}
\phi = M_\P\sqrt{\frac{3}{2}}  \ln \Phi,
\end{align}
we obtain the action in the Einstein frame
\begin{align}
	S = \int\df^{4}x \sqrt{-g_\E} \left[  \frac{M_\P^{2}}{2}R_\E  -\frac{1}{2} \pn{\partial \phi}_\E^2
	- \frac{1}{2}e^{-\sqrt{2\ov3}{\phi\ov M_\P}} \pn{\p\chi}_\E^2 - W\fn{\phi,\chi} \right], \label{Eq:EinsteinActionApp}
\end{align}
where
\begin{align}
	W(\phi,\chi) := e^{-2\sqrt{2\ov3}{\phi\ov M_\P}} \left[ V \fn{e^{\sqrt{\frac{2}{3}}{\phi\ov M_\P}},\chi } + U\fn{\chi}\right].
		\label{W given}
\end{align}
Given $f_{,\Psi\Psi}\fn{\Psi,\chi}\neq0$, the theory~\eqref{starting action} turns out to be classically equivalent to the Einstein-Hilbert gravity plus two scalars, namely the Higgs and \emph{scalaron}.
We note that there is no frame in which (or field redefinition with which) both of them are canonically normalized for all their field values.

Let us briefly see how the scalaron degree of freedom emerges by counting the number of degrees of freedom.
Pure Einstein gravity has 2 physical degrees of freedom, corresponding to gravitational waves in linearized GR: The symmetric metric tensor $ g_{\mu\nu} $ has 10 d.o.f.s, while we have 4 diffeomorphism invariance (which implies the Bianchi identity $ \nabla^{\mu} G_{\mu\nu} = 0 $) and we are left with 6. Among 10 Einstein equations of $ G_{\mu\nu} = T_{\mu\nu}/ M_\P^2 $ in pure Einstein gravity, $ 00 $-components and $ 0i $-components are actually constraints, in the sense that there is no time-derivative term in the equation of motion. These 4 constraints result in $6-4=2$ graviton degrees of freedom in pure Einstein gravity. However, by having $ R^{n} $ ($ n \neq 1 $) term, one linear combination of these equations obtains the time derivative term, which makes one scalar degree of freedom dynamical and this corresponds to the scalaron.

\subsection{Higgs-$ R^{2} $ Inflation}
By taking into account all the terms up to dimension four (i.e.\ dimension two in $ f\fn{R,\chi} $), we have
\begin{align}
	f\fn{R,\chi} &= \left(1 +	{\xi\ov M_\P^2} \chi^{2}
		+\cdots
	\right)R + \pn{{\alpha\ov M_\P^2}+\cdots} R^{2}
	+\cdots, \nn
U(\chi) &= \frac{m^{2}}{2} \chi^{2} + \frac{\lambda}{4} \chi^{4}+\cdots,
	\label{restricted f and U}
\end{align}
where the dots denote the dimension-six terms and higher that we neglect in this subsection, $ \xi $~is the non-minimal coupling, $ \alpha $ is related to the inflaton mass as we will see, $ m $ is the mass of $ \chi $, and $\lambda$ is its quartic coupling.\footnote{In principle, one may also add either $R_{\mu\nu}R^{\mu\nu}$ or $R_{\mu\nu\rho\sigma}R^{\mu\nu\rho\sigma}$ (barring the topological Gauss-Bonnet term), which will be an interesting generalization to our analysis. See also footnote~\ref{footnote:GB term}.}

This is nothing but the Higgs-$ R^{2} $ inflation model, which may be regarded as a generalization of either Higgs inflation or the $ R^{2} $ inflation model. As mentioned above, adding $ R^{2} $ term to the vanilla SM Higgs inflation cures its unitarity problem during the (p)reheating, and phenomenology of this model has been extensively studied~\cite{Ema:2017rqn,Gorbunov:2018llf,He:2018gyf,He:2018mgb,Gundhi:2018wyz,Ema:2019fdd,Cheong:2019vzl,He:2020ivk,Ema:2020evi,He:2020qcb,Lee:2021dgi,Lee:2021rzy,Panda:2022esd,Cheong:2022gfc}.

In the SM Higgs inflation, the mass of the Higgs is at the electroweak scale, which is negligible compared to the field values during the inflation. If we regard $\chi$ as a general scalar, the mass could be large so that there exists some parameter regime where the mass term becomes relevant; see Appendix~\ref{App:mass} for more discussion.

In the remainder of this section, we will first review the Higgs-$ R^{2} $ model and relevant procedures such as integrating out one degree of freedom using the valley equation and deriving an effective single-field potential. We will use the same methodology to analyze the generalizations of this model with higher dimensional operators as well.

By putting Eq.~\eqref{restricted f and U} into Eq.~\eqref{W given} via Eq.~\eqref{V given}, we obtain
\begin{align}
W\fn{\phi,\chi}
	&=	e^{- 2 \sqrt{2\ov3}{\phi\ov M_\P}} \left[ {3M_\phi^2 M_\P^2 \ov4} \left(e^{\sqrt{2\ov3}{\phi\ov M_\P}} - 1 - {\chi^{2}\ov M_\xi^2} \right)^{2} +  \frac{1}{2}m^{2}\chi^{2} + \frac{\lambda}{4} \chi^{4} \right],\label{Eq:Potential}
\end{align}
where we have introduced the scale
\al{
M_\xi
	&:=	{M_\P\ov\sqrt\xi}
}
and have interpreted
\begin{align}
M_\phi^2
	&:=	\left. \frac{\partial^{2} W}{\partial \phi^{2}} \right\vert_{\phi = \chi = 0} = \frac{ M_\P^2}{6 \alpha}
\end{align}
as the mass-squared of the $\phi$ degree of freedom at the origin of the field space. Here and hereafter, we trade $\alpha$ for $M_\phi$, and use $\xi$ and $M_\xi$ interchangeably.
To summarize, the truncation~\eqref{restricted f and U} gives the Jordan-frame action
\al{
S	&=	\int\df^4x\sqrt{-g}\sqbr{{M_\P^2\ov2}\br{\pn{1+{\chi^2\ov M_\xi^2}}R+{1\ov6M_\phi^2} R^2}-{1\ov2}\pn{\p\chi}^2-{m^2\ov2}\chi^2-{\lambda\ov4}\chi^4},
}
and the resultant Einstein-frame action reads
\al{
S	&=	\int\df^4x\sqrt{-g_\E}\sqbr{
			{M_\P^2\ov2}R_\E-{1\ov2}\pn{\p\phi}_\E^{2}-{e^{-\sqrt{2\ov3}{\phi\ov M_\P}}\ov2}\pn{\p\chi}_\E^2-W\fn{\phi,\chi}
			},
}
with $W\fn{\phi,\chi}$ being given by Eq.~\eqref{Eq:Potential}.

Now
\al{
\Phi
	&=	1+{\chi^2\ov M_\xi^2}+{R\ov3M_\phi^2},&
\phi
	&=	M_\P\sqrt{3\ov2}\ln\fn{1+{\chi^2\ov M_\xi^2}+{R\ov3M_\phi^2}}.
}
In the limit $\xi\to0$ ($M_\xi\to\infty$), $\phi$ becomes a pure \emph{scalaron} originating from the $R^2$ term, and the model reduces to the $R^2$ inflation, while in the limits $\alpha\to0$ ($M_\phi\to\infty$), one degree of freedom gets very heavy and  the $\phi$ field becomes almost Higgs $\chi$, in which the model reduces to the Higgs inflation.\footnote{
In the literature, the $\Phi$ (or $\phi$) field itself is sometimes called scalaron. Here we make the distinction between $\Phi$ (or $\phi$) and the scalaron originating from $R^2$, in order to take into account the Higgs inflation limit as well.
} As a result, for a small non-minimal coupling with $ \xi^{2}\ll\lambda{M_\P^{2}\ov M_\phi^2}$ ($M_\xi^4\gg {M_\phi^2M_\P^2 / \lambda}$), this reduces to Starobinsky inflation and we call these parameters as `$ R^{2} $-like' regime. On the other hand in the $ \xi^{2} \gg  \lambda{M_\P^{2} / M_\phi^2}$ ($M_\xi^4\ll{M_\phi^2M_\P^2 / \lambda}$) limit, we have the same result as single field inflation with the non-minimal coupling, like Higgs inflation. This is called the `Higgs-like' regime \cite{Ema:2017rqn,He:2018gyf}.

One of the feature usually overlooked in the literature is the effect of the mass term of the Higgs scalar $\chi$ field. When $m$ is non-negligible and larger than a critical mass $m_{c}$ which is a few orders of magnitude smaller than the Planck mass, the effective potential starts to develop a bump. With violation of slow-roll approximation, it may change the predictions of the model. However, it turns out that the predictions rarely change when $m $ is sufficiently small. See the Appendix~\ref{App:mass} for more detail and quantitative criterion. In most of the analyses below, we will assume $m \ll m_{c}$, and practically set $m = 0$. Then in order to have a correct amplitude of the scalar power spectrum $ A_{s} \simeq 2.1 \times 10^{-9} $, we need $ \widetilde{M}_\phi \simeq 1.3 \times 10^{-5} M_\P$; see Eq.~\eqref{M phi tilde} below for its definition. (The definition of $A_{s}$ is also given in Eq.~\eqref{eq:spectrum} below.)

\begin{figure} \centering
	\includegraphics[width=0.45\textwidth]{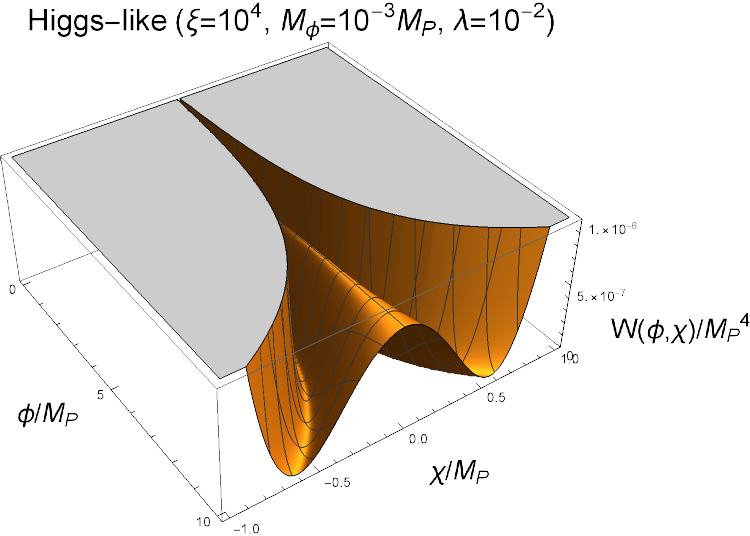}
	\includegraphics[width=0.45\textwidth]{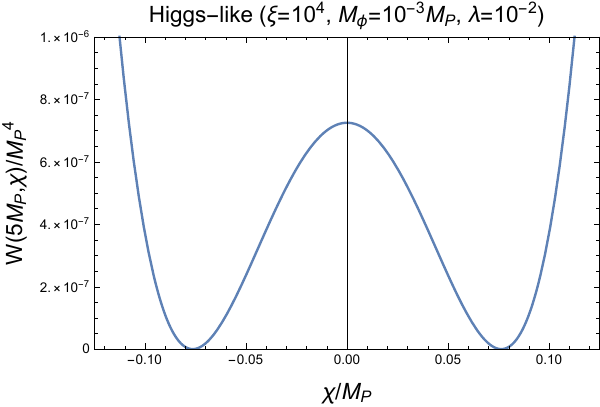}
	\caption{(Left) Shape of the potential for Higgs-like parameter ($\xi = 10^{4}, M_{\phi}=10^{-3}M_{\rm P}, \lambda = 10^{-2}$). (Right) A cross-sectional shape of the potential taking $\phi = 5 M_{\rm P}$. \label{fig:potentialHiggs}}
\end{figure}

\begin{figure} \centering
	\includegraphics[width=0.45\textwidth]{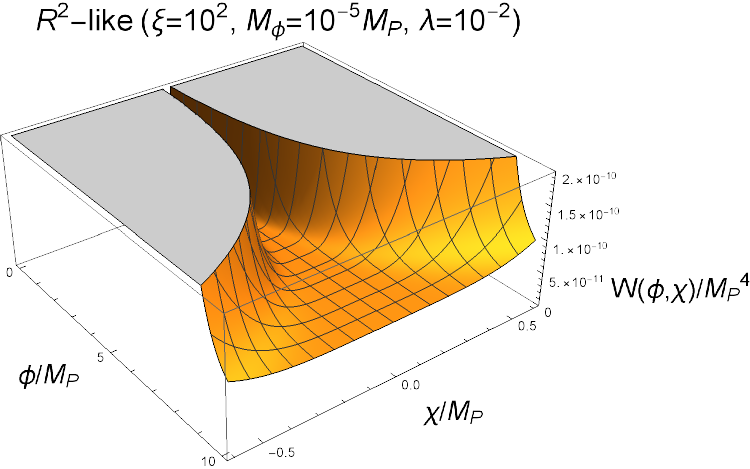}
	\includegraphics[width=0.45\textwidth]{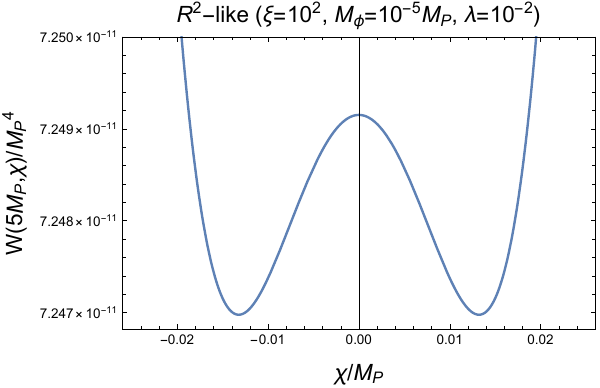}
	\caption{(Left) Shape of the potential for $R^{2}$-like parameter ($\xi = 10^{2}, M_{\phi}=10^{-5}M_{\rm P}, \lambda = 10^{-2}$). (Right) A cross-sectional shape of the potential taking $\phi = 5 M_{\rm P}$. We magnified the vertical axis to explicitly see the existence of the valley.
		\label{fig:potentialR2}}
\end{figure}

One good property of this model is that, for the majority of parameters of the model, the field trajectory during the inflation quickly becomes insensitive to the initial field value and then follows a specific trajectory along a valley of the potential; see Figure~\ref{fig:potentialHiggs} and Figure~\ref{fig:potentialR2}.
We can find the value of $\chi$ at the bottom of the valley by solving 
\al{
{\p W\ov\p\chi}
	&=	0
}
for $\chi$ in terms of $\phi$. This is justified when $\ab{\p^2W\ov\p\phi^2}\ll H_\tx{inf}^2={W\ov3M_\P^2}\ll e^{\sqrt{2\ov3}{\phi\ov M_\P}}{\p^2W\ov\p\chi^2}$ with ${\p^2W\ov\p\chi^2}>0$ along the valley, where the exponential factor is from the non-canonically normalized kinetic term for $\chi$. We have checked that this condition is safely satisfied for the parameter space $\xi\gtrsim 10$ that we have explored.

In Higgs-$ R^{2} $ inflation, following this valley, $ \chi_{\tx{v} ,0} $ can be represented as a function of $ \phi (> 0) $ as
\begin{align}
	\chi_\tx{v,0} = \pm M_\xi\sqrt{\frac{ 3{ M_\phi^{2}M_\P^2 } \left(	e^{\sqrt{2\ov3}{\phi\ov M_\P}} - 1	\right)
		}{\lambda M_\xi^4 + 3{M_\phi^2M_\P^2}}}
		\label{chi on valley}
\end{align}
and the effective single-field potential is obtained by substituting $ \chi = \chi_\tx{v,0}(\phi) $:
\begin{align}
W_{\rm eff}(\phi)
	=	\frac{3 \lambda M_\phi^{2} M_\P^2 }{4\pn{\lambda+ 3 {M_\phi^2M_\P^2\ov M_\xi^4}}}
			\left( 1 - e^{-\sqrt{2\ov3}{\phi\ov M_\P} } \right)^{2}.
\end{align}
Indeed this is 
the well-known Starobinsky-type potential that is exponentially flat as $ \phi \rightarrow \infty $ when the mass parameters are recasted into
\begin{align}
	\widetilde{M}_\phi^{2} := \frac{M_\phi^2}{1 + 3\frac{M_\phi^2M_\P^2}{\lambda M_\xi^{4}}},
		\label{M phi tilde}
\end{align}
such that the effective single-field potential is rewritten as 
\begin{align}
\label{eq:Weff_nomass}
W_{\rm eff}(\phi)
	=	\frac{3 M_\P^2 \widetilde{M}_\phi^2}{4} \left( 1- e^{-\sqrt{2\ov3}{\phi\ov M_\P} } \right)^{2} \,.
\end{align}
We note that putting the valley value~\eqref{chi on valley} into the kinetic term of $\chi$ induces additional kinetic term $-{\wt M_\phi^2\ov 4\lambda M_\xi^2}\pn{\p\phi}_\E^2$ at large $\phi \gg M_{\rm P}$,
which will be negligible in the following numerical analyses since the coefficient is very small for our choices of parameters that fit the observed value of the amplitude of the scalar power spectrum.

The scalar power spectrum that originated during inflation is typically parameterized as follows:
\begin{align}
	\mathcal{P}_{\mathcal{S}} (k) = A_{s} \left( \frac{k}{k_{*}}\right)^{n_{s}-1} \label{eq:spectrum}
\end{align}
where $A_{s}$ is the scalar amplitude, $n_{s}$ is the spectral index, and $ k_{*} $ is a pivot scale. Similarly, tensor power spectrum is parameterized with an amplitude $A_{t}$ and the tensor-to-scalar ratio is defined by $  r \equiv A_{t}/A_{s} $. Based on the Planck+BICEP/Keck 2018 result, we have $ A_{s} \simeq 2.1\times 10^{-9} $ and $ r < 0.035  $ with $ k_{*} = 0.05 ~\text{Mpc}^{-1}$ \cite{Planck:2018jri,BICEP:2021xfz}. Even though the allowed range of $n_{s}$ depends on $r$, we take $ 0.958 < n_{s} < 0.975$ from $ 1\sigma $ bound given in Ref.~\cite{BICEP:2021xfz} for later plotting purposes.

From this effective potential, slow-roll parameters are defined as
\begin{align}
	\epsilon = \frac{M_\P^2}{2} \left( \frac{W_{\rm eff}^{\prime}(\phi)}{W_{\rm eff}(\phi)}	\right)^{2}, && \eta =  M_\P^2\frac{W_{\rm eff}^{\prime\prime}(\phi)}{W_{\rm eff}(\phi)},
\end{align}
and cosmological observables such as the spectral index $n_s$, the tensor-to-scalar ratio $r$ and the amplitude of the scalar power spectrum $A_s$ defined above are evaluated as
\begin{align}
	n_{s} = 1 - 6 \epsilon + 2\eta \vert_{\phi=\phi_{*}}, && r = 16 \epsilon \vert_{\phi=\phi_{*}}, && A_{s} = \left. \frac{W_{\rm eff}}{24\pi^{2} \epsilon M_\P^4} \right\vert_{\phi=\phi_{*}},
\end{align}
at the specific field value $\phi_{*}$ of the corresponding pivot scale. In our work, the $e$-folding number is fixed to be 60 for the illustration.

This also shows unique characteristics of Higgs-$ R^{2} $ inflation. When we set $m=0$ so that the effective single-field potential is given as Eq.~\eqref{eq:Weff_nomass}, the predictions for $n_s$ and $r$ do not depend on the overall coefficient of the potential $\widetilde{M}_\phi^2 M_\P^2$, since they are canceled in the definitions of the slow-roll parameters, and hence all the parameter space for $M_\phi$, $\xi$, and $\lambda$ predicts the same $ n_{s} $ and $ r $ values.

The predictions are indeed sensitive to the addition of higher dimensional operators. We investigate how sensitive the predictions for $n_s$ and $r$ are depending on the existence of higher dimensional operators, which is the main topic in the next section.

\section{Higher dimensional operators} \label{Sec3}

In the EFT framework, we must have higher dimensional operators suppressed by a cut-off scale $ \Lambda $, which become irrelevant at low energies and small field values. However, during inflation, many models of inflation have high energy and large field values, and we have to worry about the sensitivity of the model to such an irrelevant operator. As a first step to consider the effects of these corrections, we will study specific forms of the dimension-six term for the Higgs-$ R^{2} $ model. In our model with $ R $ and $ \chi $, we have four possibilities of having dimension-six terms:\footnote{
		At dimension 4,
		there could be $ R_{\mu\nu}R^{\mu\nu} $ and $R_{\mu\nu\rho\sigma}R^{\mu\nu\rho\sigma}$ terms.
		Especially the combination of Gauss-Bonnet term
		$R^{2} - 4  R_{\mu\nu}R^{\mu\nu} + R_{\mu\nu\rho\sigma}R^{\mu\nu\rho\sigma}$ gives a total derivative and does not induce any dynamics, namely, does not affect the equation of motion at the perturbative level. At dimension 6, 
	the $\chi$ scalar may non-minimally couple to Gauss-Bonnet. While these terms do not add a new scalar degree of freedom,
	they affect to Friedman equation and potentially change the observables \cite{Satoh:2007gn,Guo:2009uk,Jiang:2013gza,Koh:2014bka,Kanti:2015pda,vandeBruck:2017voa,Nojiri:2017ncd,Odintsov:2018zhw,Nojiri:2019dwl,Odintsov:2020xji,Kawai:2021edk}. \label{footnote:GB term}}
\begin{align}
	\chi^{6}, && \chi^{4} R, && \chi^{2} R^{2}, && R^{3}.
\end{align}

However, it is not clear what $ \Lambda $ value should be taken in the EFT expansion. Three plausible possibilities are to set either $ \Lambda = M_\P $, the Planck mass; $ \Lambda = M_\phi $, the inflaton mass; or $\Lambda=M_\xi$, the non-minimal scale.

The first possibility of setting $ \Lambda = M_\P $ can be justified in the sense that it is the cut-off scale given by the unitarity of the theory. In this case, as we will see, only $ \chi^{6} $ operator gives different inflationary predictions with $ \mathcal{O}(1) $ coefficient, and inflation dynamics and predictions hardly change under the addition of the other 3 terms. Indeed, it has been shown that the system with two hierarchical scales $M_\phi\lesssim M_\xi\ll M_\P$ can be perturbative up to $M_\P$~\cite{Ema:2017rqn,Ema:2020evi}. On the other hand, since $M_\phi\to0$ does not enhance any symmetry in the action, an EFT point of view motivates to take the lowest mass scale $M_\phi$ or $M_\xi$ to be the cutoff, as in the following second and third possibilities.

The second possibility of setting $ \Lambda = M_\phi$ is adopted in many works considering higher dimension operators in the pure $ R^{2} $ inflation case. In this case, $ R^{2} $ inflation is sensitive to the addition of $ \frac{c_3}{6(3M_\phi^2)^{2}} R^{3} $ in $ f(R) $ as along as $ c_3 > \mathcal{O}(10^{-3}) $, which is sometimes used to show the sensitivity of $ R^2 $ inflation  under the presence of any higher-order corrections \cite{Huang:2013hsb,Cheong:2020rao,Ivanov:2021chn}. The fact mentioned above (namely the perturbativity up to $M_\P$) can be rephrased in this point of view as follows: If we set the coupling $\alpha$ many orders of magnitude larger than unity $\alpha\gg1$ (namely $M_\phi\ll M_\P$) and if we set all the couplings $c_n$ in the terms ${c_n\ov M_\phi^{2n-2}}R^n$ in $f\fn{R}$ ($n\geq3$) zero (or many orders of magnitude smaller than unity $c_n\ll 1$) by hand, then the cut-off scale of the $ R^{2} $ inflation becomes also $M_\P$~\cite{Hertzberg:2010dc,Kehagias:2013mya} because the scalaron with the small mass $M_\phi$ becomes a dynamical degree of freedom in the theory that cures the unitarity up to $M_\P$.

The third possibility of setting $\Lambda=M_\xi$ changes the inflationary predictions when a coupling constant of a non-minimal term such as $\chi^4R$ and $\chi^2R^2$ becomes of order unity as we will see. Again, the unitarity would be restored up to $M_\P$ if we set the coupling $\alpha$ many orders of magnitude larger than unity (namely $M_\phi\ll M_\P$) and if we set all the higher dimensional couplings to zero (or many orders of magnitude smaller than unity) by hand.

To transparently see these observations, we parameterize the dimension-six terms and their dimensionless coefficients $\beta_1$, $\beta_2$, and $\beta_3$ as follows:
\begin{align}
f\fn{R,\chi}
	=	\left(1 + \frac{\chi^{2}}{M_\xi^{2}}+\beta_{3}{\chi^{4}\ov M_\xi^4}+\cdots\right)R
		+\pn{{1\ov6}+\beta_2{\chi^2\ov M_\xi^2}+\cdots}\frac{R^2}{M_\phi^2}
		+\Pn{\beta_{1}+\cdots}\frac{R^{3}}{M_\phi^{4}}
		+\cdots,
		\label{our parametrization}
\end{align}
where the dots denote dimension-eight terms and higher.
For the Jordan-frame Higgs potential, we set $ \Lambda = M_\P $:
\begin{align}
	U(\chi) = \frac{\lambda}{4} \chi^{4} + \gamma\frac{\chi^{6}}{M_\P^{2}} .
		\label{potential with sixth}
\end{align}

For simplicity, we will turn on each of the couplings $\beta_1$, $\beta_2$, $\beta_3$, and $ \gamma $, one by one, and see how the required values of $ M_\phi$ and $ \xi  (=M_\P^2/M_\xi^2$) to obtain $A_{s}  \simeq 2.1 \times 10^{-9}$ are altered, 
and how the predictions for $ n_{s}$ and $r$ change accordingly. For each case, we will present how the potential shape changes in both $ R^{2} $-like and Higgs-like parameters, and how predictions change for $ n_{s} $ and $ r $.

\subsection{$ \chi^{6} $ term}

\begin{figure} [t] \centering
	\includegraphics[width=0.49\textwidth]{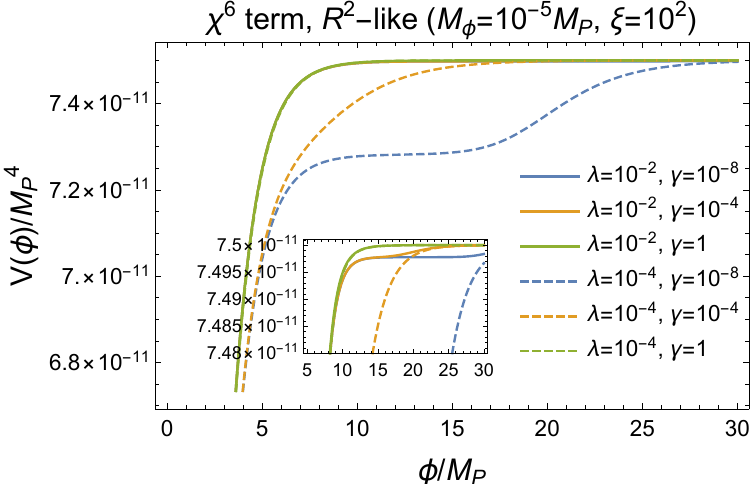}
	\includegraphics[width=0.45\textwidth]{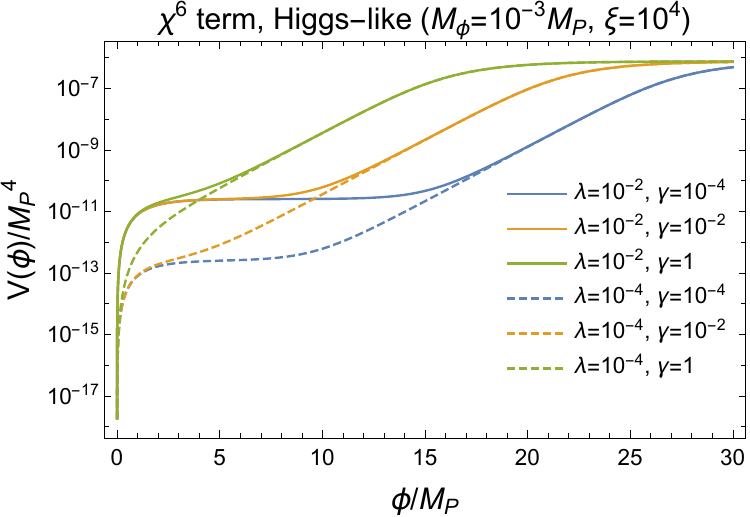}
	\caption{\label{Fig:chi6potential} Change of the potential shape for the $ R^{2} $-like ($ M_\phi = 10^{-5} M_{\rm P} $, $ \xi = 10^{2} $) and Higgs-like ($ M_\phi = 10^{-3} M_{\rm P} $, $ \xi= 10^{4} $) for various $ \gamma $ and $ \lambda = \left(10^{-2}, 10^{-4} \right) $, in the presence of $ \chi^{6} $ term.
	}
\end{figure}

First, we consider the $ \chi^{6} $ term in the potential with nonzero $ \gamma $, while setting all other higher dimensional operators to vanish (i.e., $ \beta_{i=1,2,3} = 0 $). With the addition of $ \chi^{6} $ term, the trajectory of the valley for the effective potential changes from the one given in Eq.~\eqref{chi on valley} as
\begin{align}
	\chi_{\text{v},\gamma} 
	 \simeq \chi_{\text{v},0} 
	- \frac{9 \sqrt{3} \gamma M_{\phi}^{3} M_{\rm P}^{3} \xi^{3/2}}{(M_{\rm P}^{2} \lambda + 3 M_{\phi}^{2} \xi^{2})^{5/2}} \left(	e^{\sqrt{\frac{2}{3}} \frac{\phi}{M_{\rm P}} }	- 1 \right)^{3/2}  + \mathcal{O}(\gamma^{2})
\end{align}
where we took small $\gamma$ limit. Also, the effective single-field potential is modified. Since its full expression is lengthy, we do not give it here, but show the shape of the effective potential for the Higgs-like regime ($ M_\phi=10^{-3}M_\P $, $ \xi = 10^{4} $) and the $ R^{2} $-like regime ($ M_\phi = 10^{-5}M_\P $, $ \xi = 10^{2} $) in Figure~\ref{Fig:chi6potential},  varying $ \lambda $ and $ \gamma $ for illustration. It is interesting that we have a transition of the potential to larger values at $ \phi \rightarrow \infty $ limit as
\begin{align}
	V(\phi) \xrightarrow{\phi \rightarrow \infty} \frac{3}{4}M_\phi^2M_\P^{2} > \frac{3}{4} M_\phi^2 M_\P^{2} \left(	 1 + 3 \frac{\xi^{2}}{\lambda} \left(\frac{M_{\phi}}{M_\P}\right)^{2}	\right)^{-1}.
\end{align}
Therefore, potential still possesses shift symmetry in $ \phi \rightarrow \infty $ limit and does not exceed the Planck scale. However, this does not successfully explain the current observation because the field value corresponding to 60 $e$-folds during inflation is below the plateau at $\phi\to\infty$ for the parameters that we scanned.

\begin{figure} [t] \centering
	\includegraphics[width=0.45\textwidth]{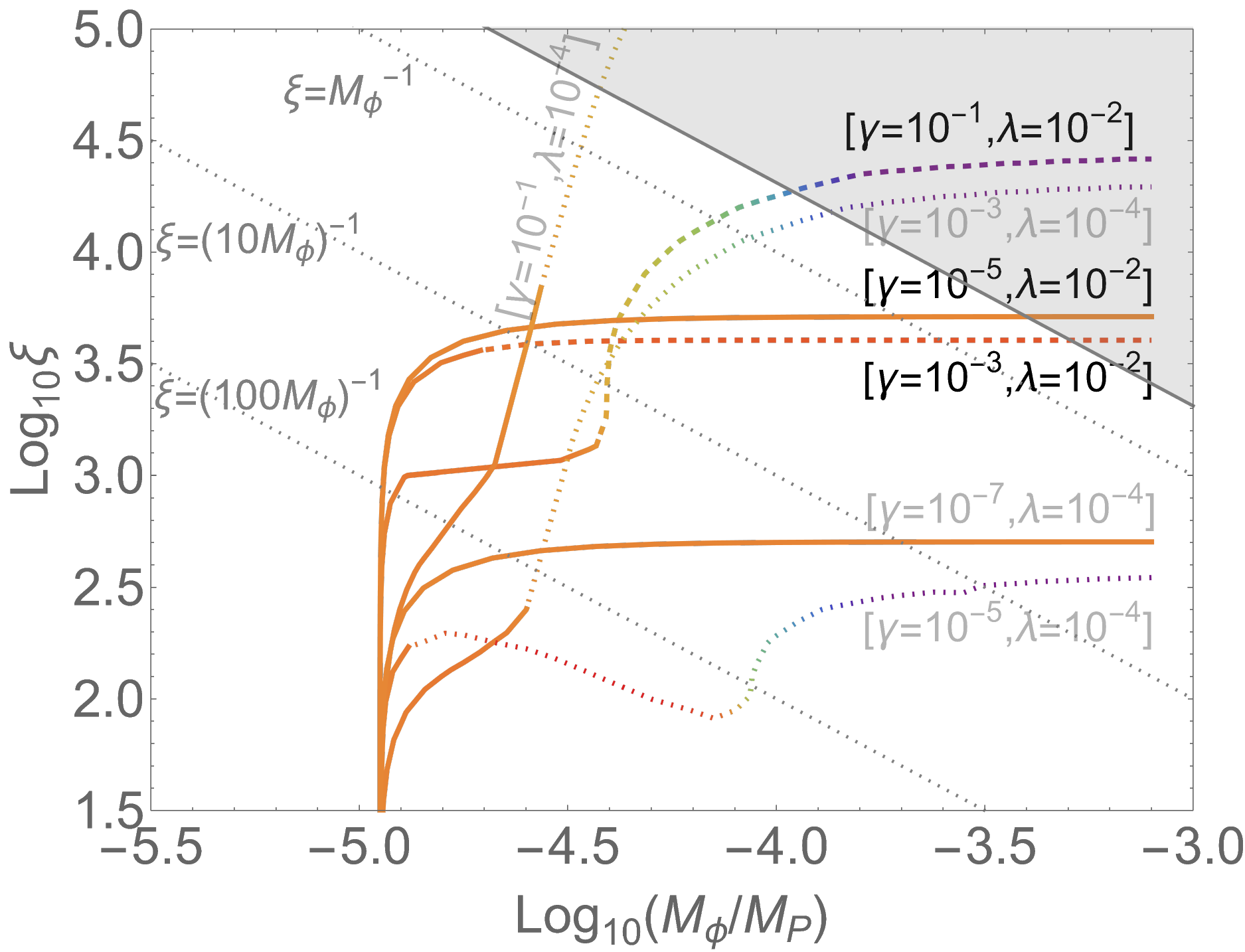}
	\includegraphics[width=0.45\textwidth]{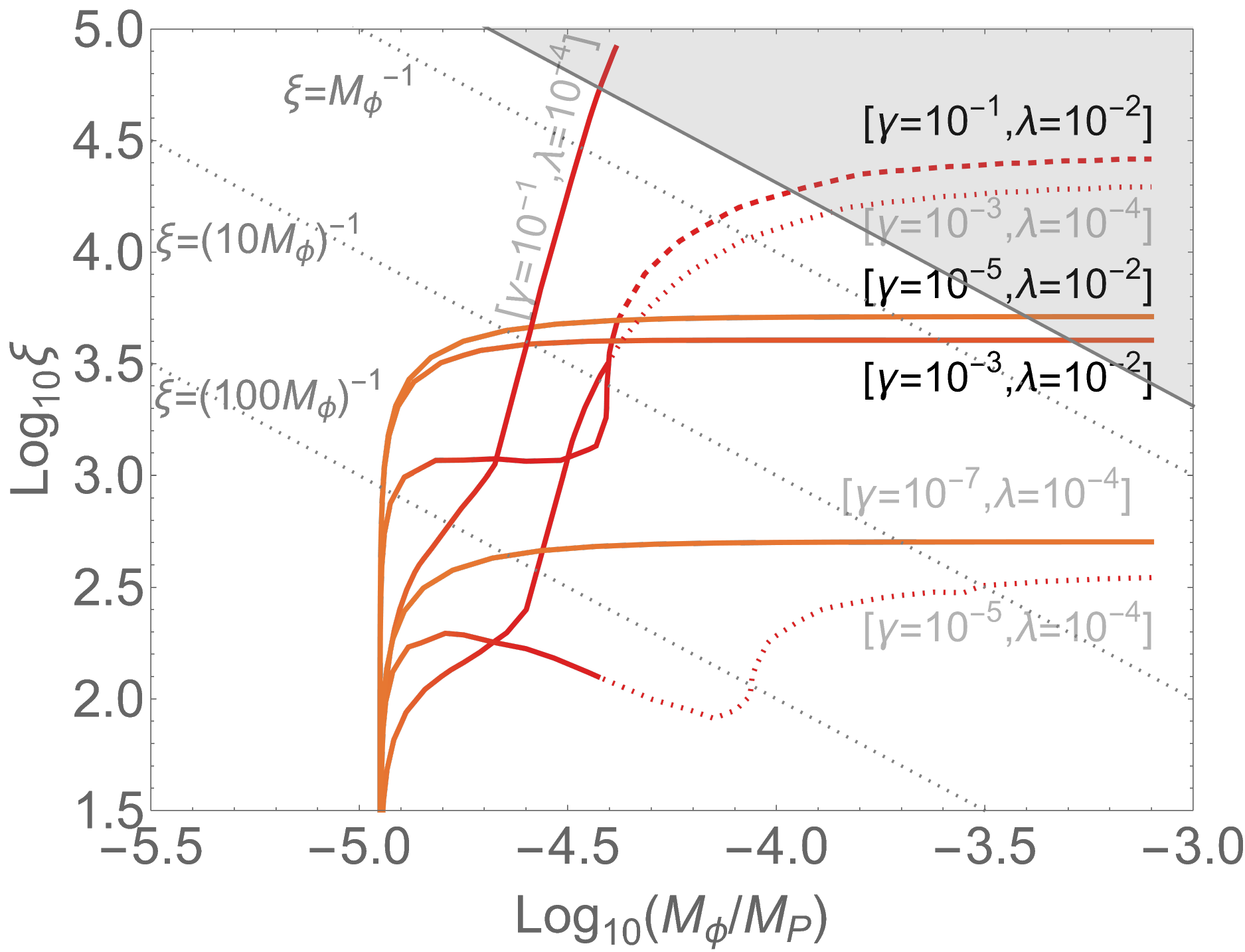}
	\caption{Deformation of cosmological observables $ n_{s} $ (left) and $ r $ (right) values in the presence of $ \chi^{6} $ term. The case $ \lambda = 10^{-2}$ ($10^{-4}$) is drawn with the colored solid line denoting allowed region and with the colored dashed (dotted) line denoting the excluded region. The straight gray dotted lines denote the values of $\xi$. The gray region is forbidden by the unitarity cutoff. 
	 \label{Fig:chi6nsr} }
\end{figure}

In Figure~\ref{Fig:chi6nsr}, we show the parameter set which fits $ A_{s} \simeq 2.1 \times 10^{-9} $ in the $ (M_\phi,\xi) $ plane, with different colors denoting the values of $ n_{s} $ and $ r $ assuming 60 $e$-folds at the pivot scale. From the figure, one can see that the so-called Higgs-like region ($ \xi^2 \gg \lambda \frac{M_P^2}{M_\phi^2}$) is more sensitive to the addition of Planck suppressed higher dimensional operator $ \chi^{6} $. Among four possible dimension-6 operators, we see only this $ \chi^{6} $ term is highly sensitive even with the Planck scale set to be the cut-off scale.  One can also find that the original predictions of purely Higgs-$ R^{2} $ model start to deviate from  $ \gamma \simeq 10^{-3} $ for $ \lambda = 10^{-2} $, and $ \gamma \simeq 10^{-6} $ for $ \lambda = 10^{-4} $. Albeit we do not consider any combinations of higher dimensional operators,  the smaller $ \lambda  $ is,  the more sensitively the predictions depend on the presence of the higher dimensional operator.

One might expect better behavior of Higgs-$ R^{2} $ inflation under the presence of $ \chi^{6} $ term than the vanilla Higgs inflation model, due to the smaller field values of Higgs in the Einstein frame $ \sim M_\P/\sqrt{\xi} $ rather than being a super-Plankian value. However, what matters is the energy density, not the field value itself. In fact, the contribution coming from $ \lambda \chi^{4} $ term in the effective potential is $ \sim (\lambda/ \xi^{2}) M_\P^{4} $ which is comparable to the energy density during the inflation in Higgs-like case. Therefore, even if the field value of the Higgs is smaller than the one in Higgs-$R^{2}$ inflation in the case with no $R^{2}$ term, $\gamma \chi^{6} / M_{\rm P}^{2}$ with $\gamma \sim 1$ is enough to change the cosmological observables.

\subsection{$ R^{3} $ term}

Next we will add $ R^{3} $ term with $ \beta_{1} $ as a perturbation. In this case, the value of $\chi$ along the trajectory of the valley is given by
\begin{align}
	\chi_{\text{v},\beta_{1}} 
& \simeq \chi_{\text{v},0} - \frac{27\sqrt{3}}{2} \beta_{1} M_{\phi} M_{\rm P}^{5} \lambda^{2} \sqrt{ \frac{    \left(e^{\sqrt{\frac{2}{3}} \frac{\phi}{M_{\rm P}}} -1 \right)^{3} \xi}{(M_{\rm P}^{2} \lambda + 3M_{\phi}^{2} \xi^{2})^{5}} }  + \mathcal{O}(\beta_{1}^{2})
\end{align}
at small $\beta_{1}$ limit.  The pure Starobinsky inflation is realized when $ \beta_1 \rightarrow 0 $ and $\xi \rightarrow 0 $.

As we discussed, it is known that, the pure $ R^{2} $ inflation is sensitive to addition on $ R^{3} $ term, once the cut-off scale of this higher dimensional operator is taken to be $ \Lambda \sim M_\phi$, while it is not much affected by Planck suppressed operators.

\begin{figure}[t] \centering
	\includegraphics[width=0.43\textwidth]{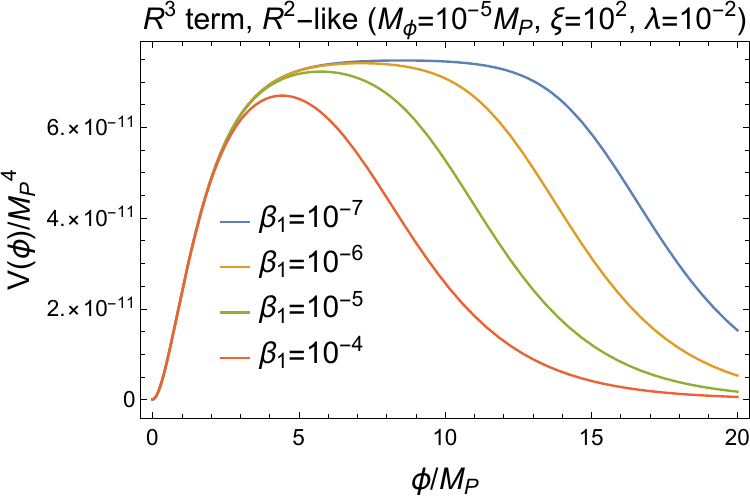}
	\includegraphics[width=0.45\textwidth]{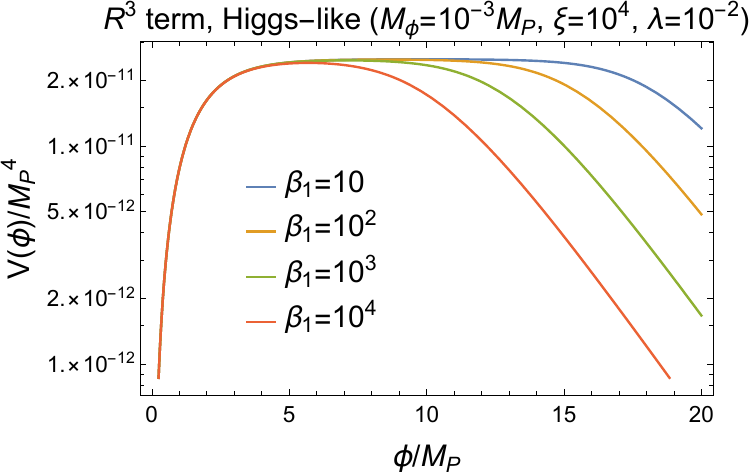}
	\caption{Change of the potential shape for the $ R^{2} $-like ($ M_{\phi} = 10^{-5} M_{\rm P} $, $ \xi = 10^{2} $, $ \lambda = 10^{-2} $) and Higgs-like ($ M_{\phi} = 10^{-3} M_{\rm P} $, $ \xi= 10^{4} $, $ \lambda = 10^{-2} $) for various $ \beta_{1} $, in the presence of $ R^{3} $ term. \label{Fig:R3_potential}}
\end{figure}

\begin{figure}[t] \centering
	\includegraphics[width=0.45\textwidth]{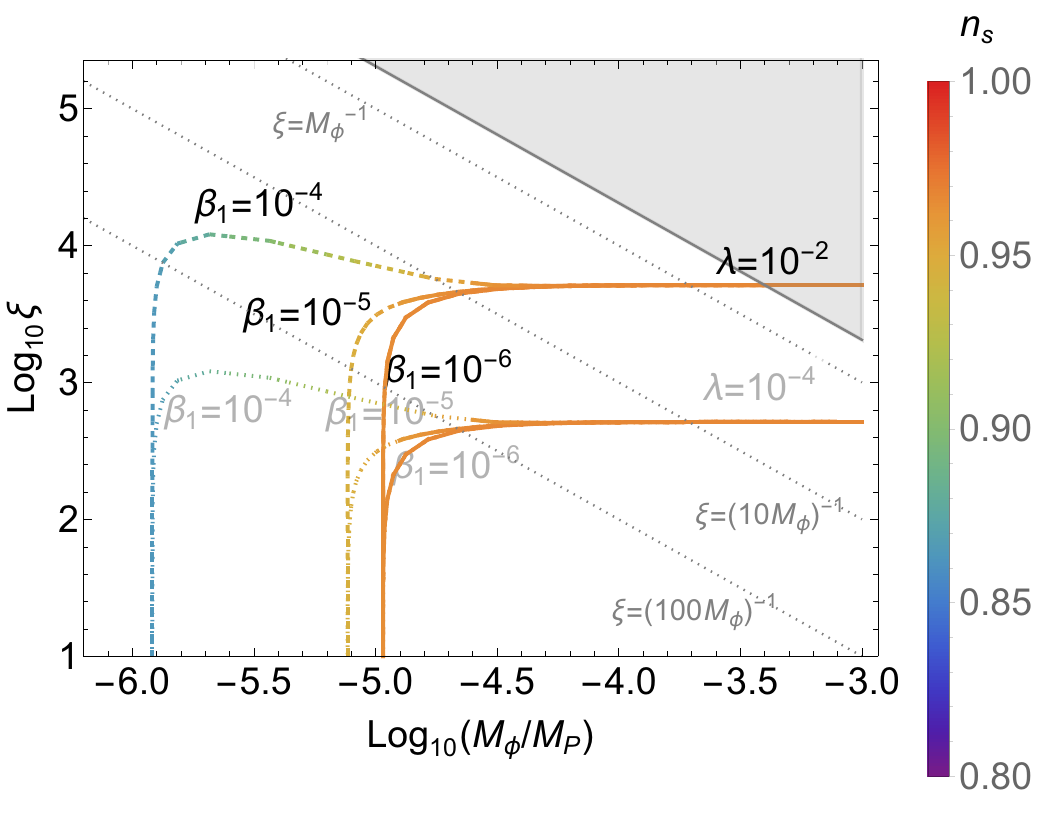}
	\includegraphics[width=0.45\textwidth]{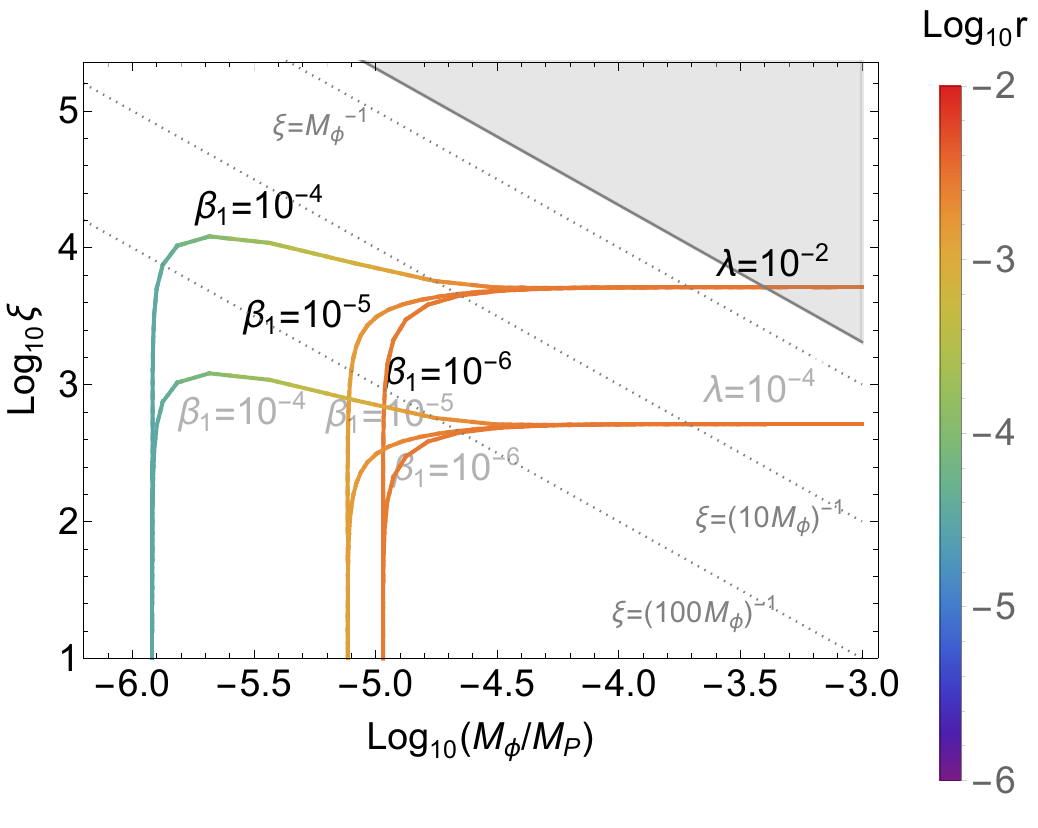}
	\caption{Deformation of cosmological observables $ n_{s} $ (left) and $ r $ (right) values in the presence of $ R^{3} $ term. The case $ \lambda = 10^{-2}$ ($10^{-4}$) is drawn with the colored solid line denoting the allowed region and with the colored dashed (dotted) line denoting the excluded region. For $ r $ plot in the right panel, all lines are allowed. The straight gray dotted lines denote the values of $\xi$. The gray region is forbidden by the unitarity cutoff.
\label{Fig:R3nsr}}
\end{figure}

This is because, from the existence of $ R^{3} $ term (more generally, any power $ n > 2 $), the potential is decreasing at a large field limit and loses its asymptotic flatness, as seen from Figure~\ref{Fig:R3_potential}. For a larger $ \beta_{1}  $, to support 60 $e$-folds, the inflaton should start near the local maximum and the field value should be fine-tuned (see also the last comment in Section~\ref{Sec:Conclusion and Discussion} for a positive use of this situation). This is basically what happens to generic small field inflation models, where the initial field value should be very close to the origin to support enough $e$-folds during the inflation \cite{GOLDWIRTH1992223,Brandenberger:2016uzh}. More quantitatively, as shown in Figure~\ref{Fig:R3nsr}, inflationary predictions on $ (n_{s},r) $ starts to change for $ \beta_{1} \geq \mathcal{O}(10^{-5}) $, as already well-known in many literature \cite{Huang:2013hsb,Cheong:2020rao}. Also, we note that predictions on $ r $ always decrease for this case and the two other following cases. Hence, the consistency to the observation mainly comes from $ n_{s} $ predictions at the moment.

On the other hand, with the large non-minimal coupling, we find that the potential shape is stable under the addition of the $ R^{3} $ term, for the parameter satisfying the current measured value of scalar power spectrum amplitude $ A_{s} \simeq 2.1 \times 10^{-9} $. However, as we will see, the Higgs-like regime also starts to drastically change once we consider other form of the higher dimension operators.\footnote{
One may wonder if large $\beta_{1}$ induce large loop corrections.
It turns out that, in the Einstein frame, all the self-couplings and coupling between $\phi$ and $\chi$ are suppressed, so we do not expect large loop corrections from the $R^{3}$ term involving large $\beta_{1}$. The same conclusion applies to other two terms $\chi^{2} R^{2}$ and $\chi^{4} R$ discussed below with $\beta_{2}$ and $\beta_{3}$, respectively.
}

Note that with our parameterization, for the same values of $ \beta_{1} $, the observables $ n_{s} $ and $ r $ only depend on the ratio $ \xi^{2}/\lambda $. Therefore, assuming two orders of magnitude smaller $ \lambda $ (corresponding to the dotted line in Figure~\ref{Fig:R3nsr}) means that one order of magnitude smaller $ \xi $ gives proper $n_s$ and $r$ as well as the scalar amplitude, which is presented by the vertical shift of the lines with the same color scheme in Figure~\ref{Fig:R3nsr}. This dependence also holds for two other cases with $ \beta_{2} $ and $ \beta_{3} $ discussed below.


We can also consider the case where the $R^3$ term is included in $f(R,\chi)$ without $R^2$, but do not discuss such a case here since it deviates from the spirit of our argument here. Nevertheless we discuss such a case in Appendix~\ref{app:R3}.
\subsection{$ \chi^{2} R^{2} $ and $ \chi^{4} R $ terms}

\begin{figure}[t]\centering
	\includegraphics[width=0.43\textwidth]{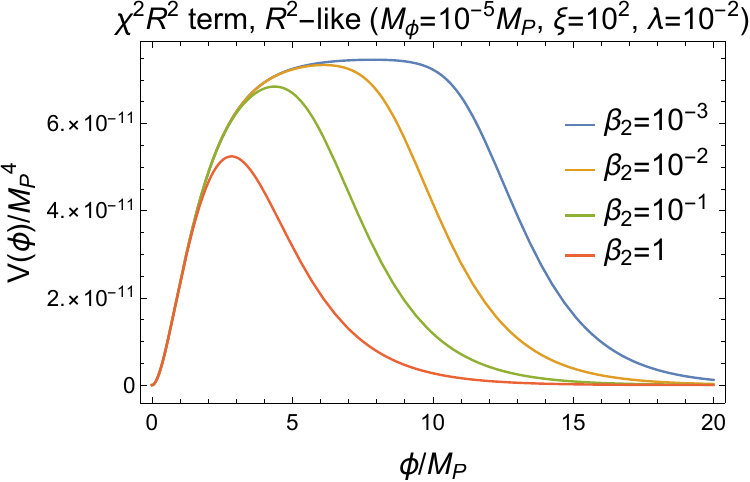}
	\includegraphics[width=0.48\textwidth]{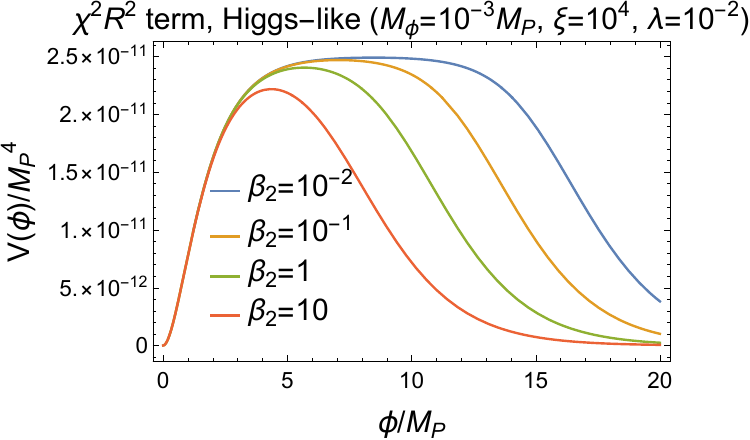}
	\caption{Change of the potential shape for the $ R^{2} $-like ($ M_{\phi} = 10^{-5} M_{\rm P} $, $ \xi = 10^{2} $, $ \lambda = 10^{-2} $) and Higgs-like ($ M_{\phi} = 10^{-3} M_{\rm P}$, $ \xi= 10^{4} $, $ \lambda = 10^{-2} $) for various $ \beta_{2} $, in the presence of $ \chi^{2} R^{2} $ term. \label{Fig:Chi2R2Potential}}
\end{figure}

\begin{figure}[t]\centering
	\includegraphics[width=0.45\textwidth]{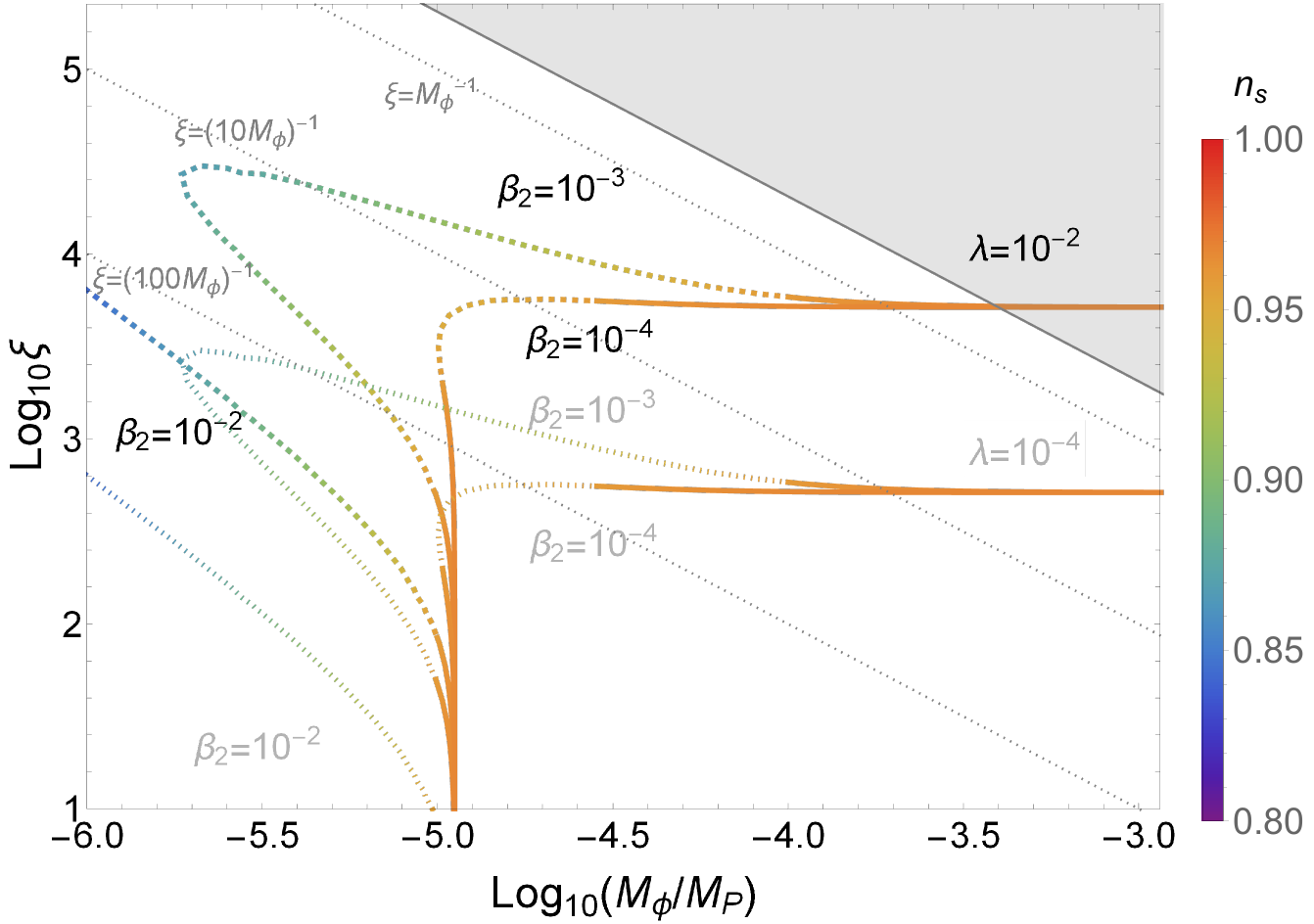}
	\includegraphics[width=0.45\textwidth]{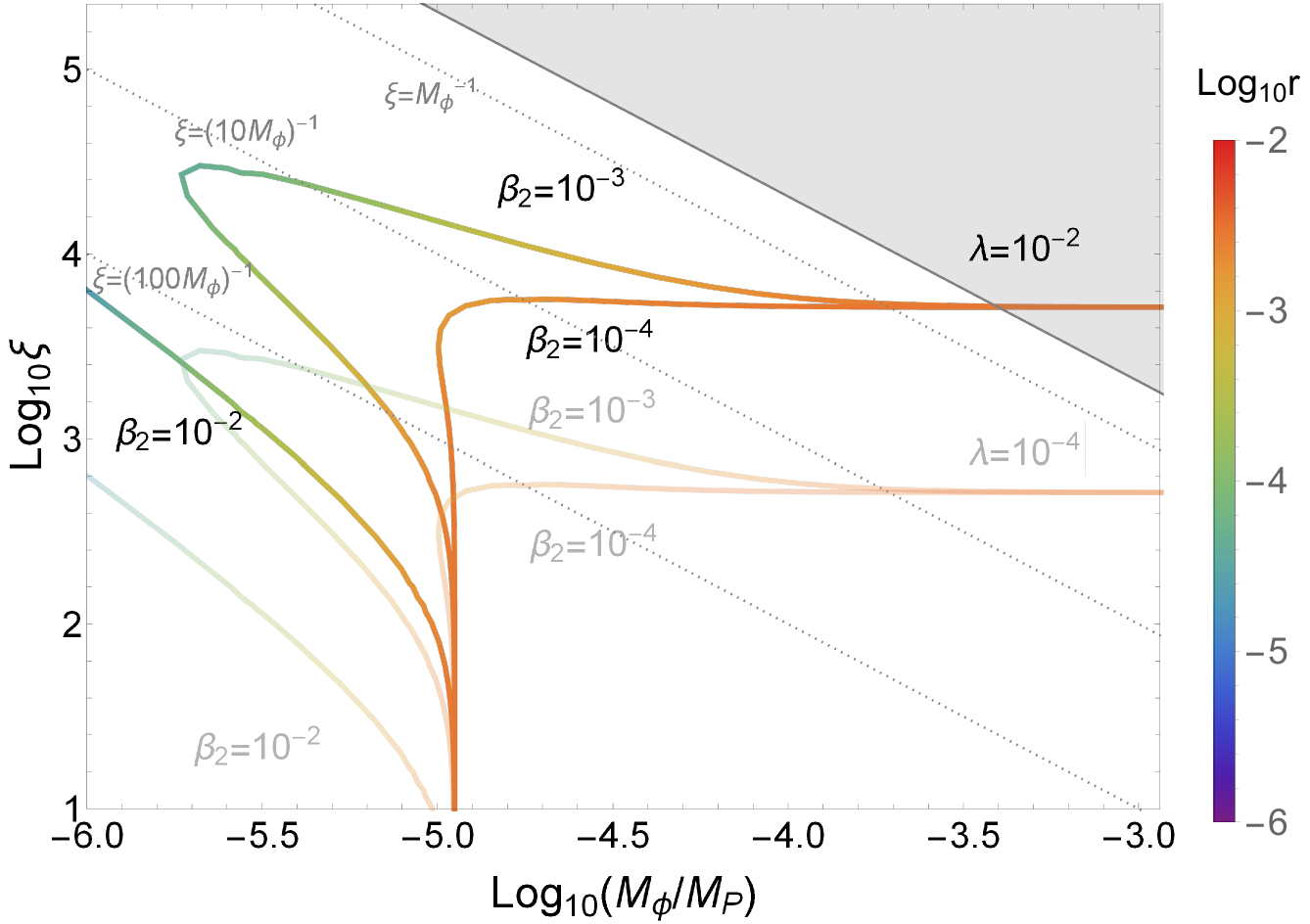}
	\caption{Deformation of cosmological observables $ n_{s} $ (left) and $ r $ (right) values in the presence of $ R^{2} \chi^{2} $ term. The case $ \lambda = 10^{-2}$ ($10^{-4}$) is drawn with the colored solid line denoting the allowed region and with the colored dashed (dotted) line denoting the excluded region. For $ r $ plot in the right panel, all lines are allowed. The straight gray dotted lines denote the values of $\xi$. The gray region is forbidden by the unitarity cutoff.
		\label{Fig:R2Chi2nsr}}
\end{figure}

\begin{figure}[t]\centering
	\includegraphics[width=0.43\textwidth]{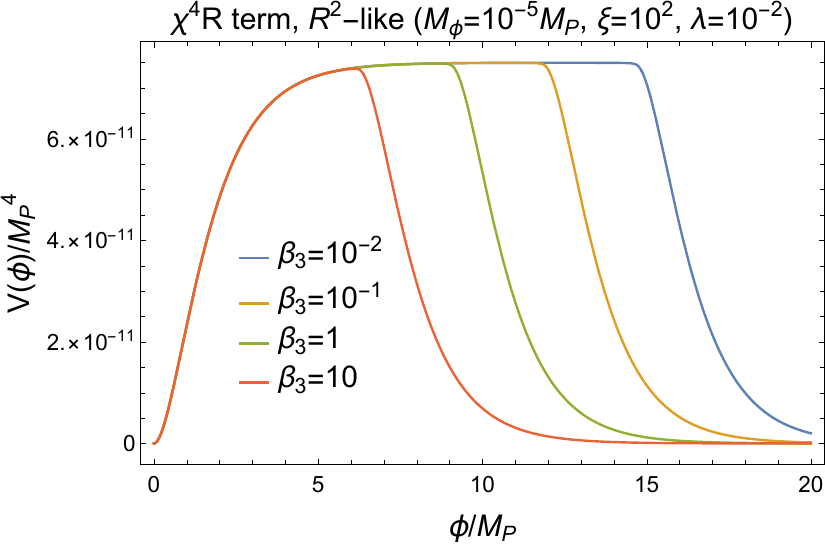}
	\includegraphics[width=0.45\textwidth]{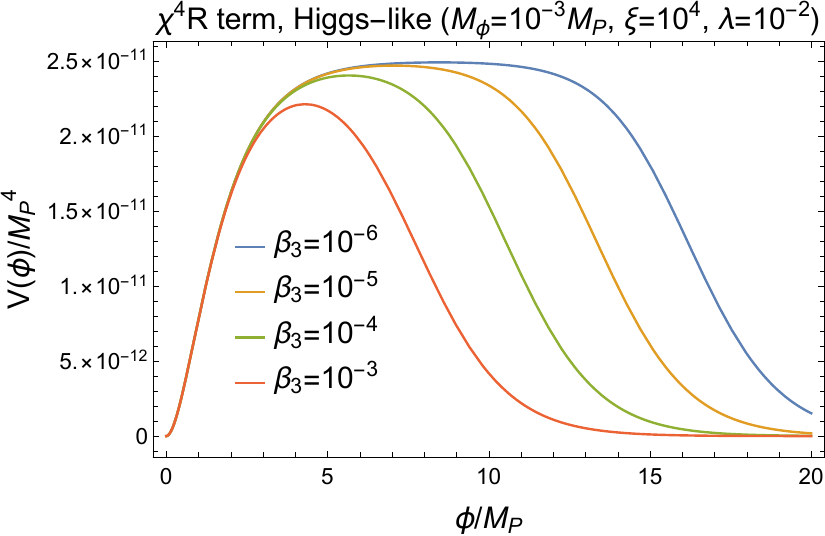}
	\caption{Change of the potential shape for the $ R^{2} $-like ($ M_{\phi} = 10^{-5} M_{\rm P} $, $ \xi = 10^{2} $, $ \lambda = 10^{-2} $) and Higgs-like ($ M_{\phi} = 10^{-3} M_{\rm P} $, $ \xi= 10^{4} $, $ \lambda = 10^{-2} $) for various $ \beta_{3} $, in the presence of $ \chi^{4}R $ term. \label{Fig:chi4RPotential}}
\end{figure}

\begin{figure}[h]\centering
	\includegraphics[width=0.45\textwidth]{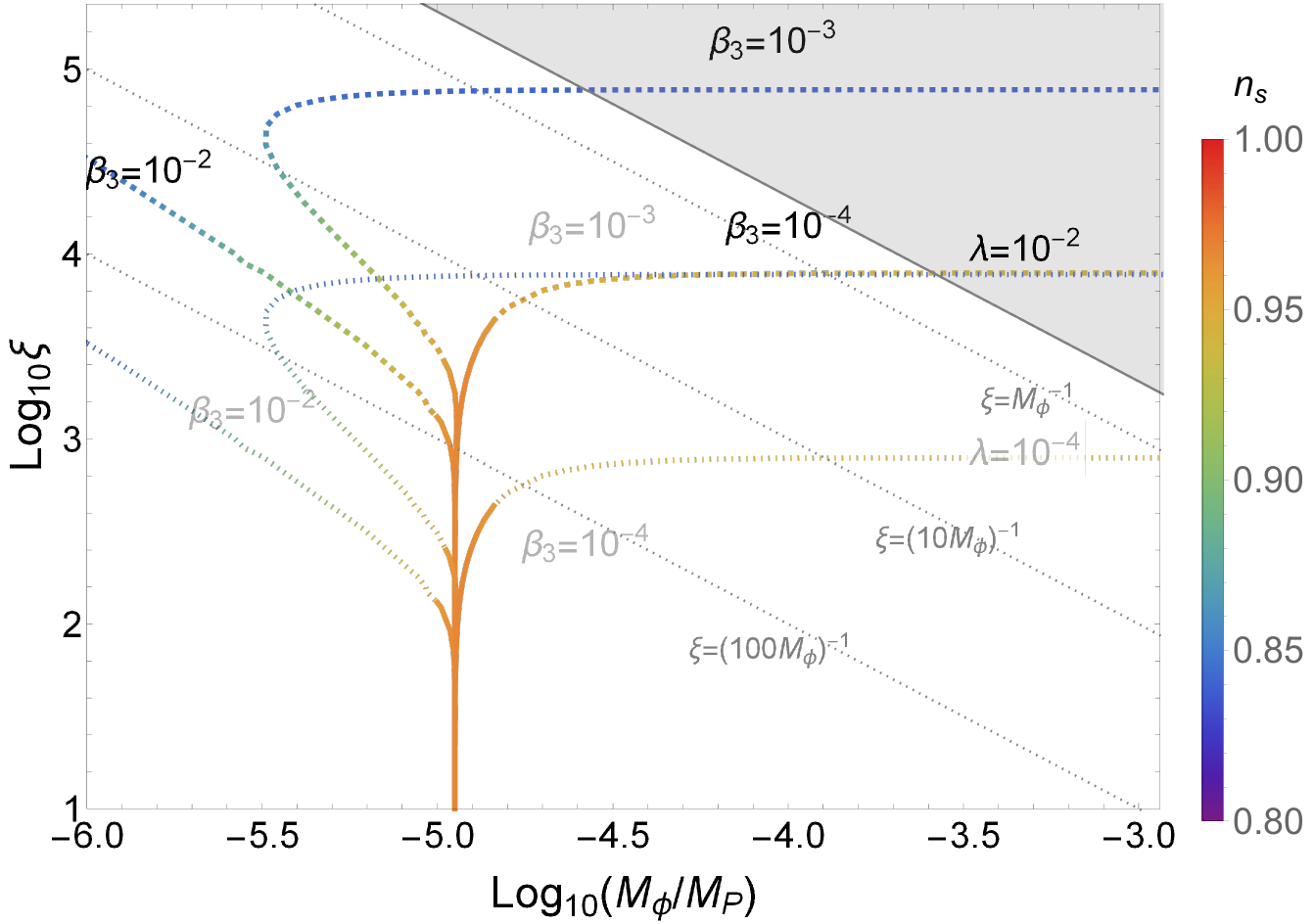}
	\includegraphics[width=0.45\textwidth]{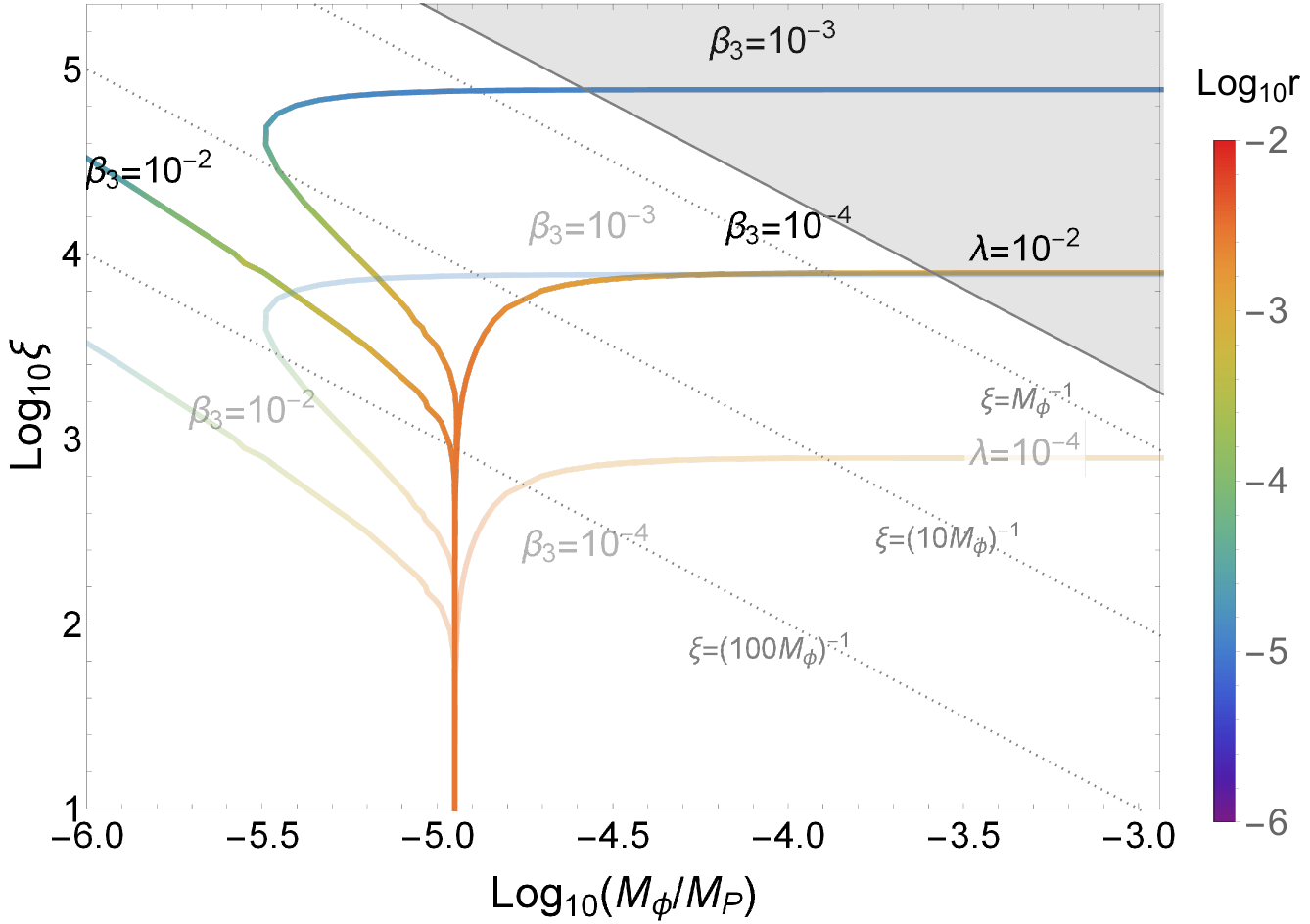}
	\caption{Deformation of cosmological observables $ n_{s} $ (left) and $ r $ (right) values in the presence of $ R \chi^{4} $ term. The case $ \lambda = 10^{-2}$ ($10^{-4}$) is drawn with the colored solid line denoting the allowed region and with the colored dashed (dotted) line denoting the excluded region. For $ r $ plot in the right panel, all lines are allowed. The straight gray dotted lines denote the values of $\xi$. The gray region is forbidden by the unitarity cutoff.
		 \label{Fig:RChi4nsr}}
\end{figure}

The deformation of potential and cosmological observables are depicted in Figure~\ref{Fig:Chi2R2Potential} and Figure~\ref{Fig:R2Chi2nsr} for $\chi^{2} R^{2}$ term, and in Figure~\ref{Fig:chi4RPotential} and Figure~\ref{Fig:RChi4nsr} for $\chi^{4} R$ term, respectively.

For $ \chi^{2} R^{2} $ case, both $R^{2}$-like and Higgs-like
regimes start to deviate as $ \beta_{2} \gtrsim \mathcal{O}(10^{-4}) $. On the other hand for $ \chi^{4} R $ case, Higgs-like regime first starts to be modified when $ \beta_{3} \gtrsim \mathcal{O}(10^{-4}) $ and $ R^{2} $-like regime is relatively stable. As we mentioned above, the values of $\beta_{2}$ and $\beta_{3}$ which start to change predictions do not depend on $\lambda$.

For these cases,   the values of $\chi$ along the valley trajectory can be written, at small $\beta_2$ and $\beta_3$ limits, 
\begin{align}
	\chi_{\text{v},\beta_{2}} & \simeq \chi_{\text{v},0} + \beta_{2} \frac{3\sqrt{3} M_{\phi} M_{\rm P}^{3} \lambda \sqrt{\xi} (M_{\rm P}^{2} \lambda - 6 M_{\phi}^{2} \xi^{2}) }{2(M_{\rm P}^{2} \lambda + 3 M_{\phi}^{2} \xi^{2})^{5/2}} \left(	e^{\sqrt{\frac{2}{3}} \frac{\phi}{M_{\rm P}}} - 1	\right)^{3/2} + \mathcal{O}(\beta_{2}^{2}),\\
		\chi_{\text{v},\beta_{3}} & \simeq \chi_{\text{v},0} + \beta_{3}
		\frac{3\sqrt{3} M_{\phi}^{3} M_{\rm P} \xi^{5/2} (2M_{\rm P}^{2} \lambda - 3 M_{\phi}^{2} \xi^{2})}{2 (M_{\rm P}^{2} \lambda + 3M_{\phi}^{2} \xi^{2})^{5/2}} 
		\left(	e^{\sqrt{\frac{2}{3}} \frac{\phi}{M_{\rm P}}} - 1	\right)^{3/2} + \mathcal{O}(\beta_{3}^{2}),
\end{align}
respectively.

Note that, from our parameterization, one may compare different conventions of choosing the cut-off scale.
For example, if one chooses purely $ M_\phi$ as a cutoff scale, the parameterization becomes
$  \frac{\widetilde{\beta}_{2}}{M_\phi^{4}} := \frac{\beta_{2}}{M_\xi^2M_\phi^2} $ and
$ 	\frac{\widetilde{\beta}_{3}}{M_\phi^{4}} := {\beta_{3}\ov M_\xi^4} $.
Therefore, for small $ \xi \sim \mathcal{O}(1) $ case, the $ \widetilde{\beta}_{2} $ coefficient should be further suppressed as 
$\widetilde{\beta_2}={M_\phi^2\ov M_\xi^2}\beta_2 
\sim {M_\phi^2 \ov M_{\rm P}^2} \beta_2 \sim 10^{-10}\beta_2$ for $R^2$-like case, implying that an EFT expansion solely with $ M_\phi$ is unnatural (or else we have to have $ \xi ={M_\text{P}^2\ov M_\xi^2}\sim \frac{M_\text{P}}{M_\phi} $, as expected in the renormalization group equation in Higgs-$ R^{2} $ inflation \cite{Ema:2019fdd}). On the other hand, if we choose $ M_\P $ as a cut-off scale of the theory, $ \frac{\widetilde{\widetilde\beta}_{2}}{M_\P^{4}} := \frac{\beta_{2}}{M_\xi^2M_\phi^2} $ so that $\widetilde{\widetilde\beta}_{2} = \frac{M_\text{P}^4}{M_\xi^2M_\phi^2} \beta_{2} $, we need a coupling multiplied by the huge factor $\frac{M_\text{P}^4}{M_\xi^2M_\phi^2}$ to affect physics, again implying the appropriateness of the parameterization~\eqref{our parametrization}.


\section{Conclusion and discussion} \label{Sec:Conclusion and Discussion}

In this work, we considered the UV sensitivity of the $ f(\chi,R) $ theory of inflation. By first considering all possible dimension-4 operators, this coincides with the Higgs-$ R^{2} $ inflation model (up to mass term), which pushes unitarity cutoff of the vanilla-Higgs inflation $ \Lambda \sim M_\P / \xi $ at the origin up to Planck scale \cite{Ema:2017rqn,Gorbunov:2018llf,He:2018gyf,He:2018mgb,Gundhi:2018wyz,Ema:2019fdd}. In this respect, one might expect this model to be insensitive to Planck-suppressed operators. While, sometimes, scalaron mass $ M_\phi$ is taken to be an EFT expansion parameter when considering higher dimension operators.

As a first step, we added all possible dimension-6 operators, term by term, to see how sensitive cosmological observables are to these modifications. To take into account the role of large non-minimal coupling lowering down the effective cut-off scale, we properly multiply factors of $ \xi $ and $ M_\phi$, with dimensionless coefficient $ \gamma $ and $ \beta_{1,2,3} $.

We have found that taking $ M_\tx{P}$ as an expansion parameter requires abnormally huge coupling for the higher dimension operators to change the cosmological observables of the original Higgs-Starobinsky model except for $\chi^6/M_\tx{P}^2$ to affect physics.
On the other hand, when $ M_\phi $ and $M_\xi$ are properly taken into account as in Eq.~\eqref{our parametrization}, the couplings $\gtrsim10^{-3}$ start to affect the predictions.\footnote{
More precisely, for the $\chi^6$ term, $\gamma/\lambda\gtrsim10^{-3}$ starts to affect the result.
}

We have seen that the slight addition of $\beta_1$, $\beta_2$, and $\beta_3$ changes the asymptotic behavior of the large field region of the potential to be a runaway one. This behavior may be used to support eternal inflation, called the topological inflation~\cite{Hamada:2014raa}, which is also suggested in string theory~\cite{Hamada:2015ria}.

\acknowledgments

We thank Yohei Ema, Koichi Hamaguchi, Kunio Kaneta, Taichiro Kugo, Shinya Matsuzaki, Seong Chan Park, and Masatoshi Yamada for inspiring discussions. SML is grateful to KIAS and CERN for its hospitality while this work was in progress. The work of SML is in part supported by the Hyundai Motor Chung Mong-Koo Foundation Scholarship, and funded by Korea-CERN Theoretical Physics Collaboration and Developing Young High-Energy Theorists fellowship program (NRF-2012K1A3A2A0105178151). The work of K.O.\ is in part supported by the JSPS KAKENHI Grant Nos.~19H01899 and 21H01107. The work of T.T.\ is in part supported by the JSPS KAKENHI Grant No.~19K03874. TM is funded by the Deutsche Forschungsgemeinschaft (DFG, German Research Foundation) under grant 396021762 - TRR 257: Particle Physics Phenomenology after the Higgs Discovery
and Germany’s Excellence Strategy EXC 2181/1 - 390900948 (the Heidelberg STRUCTURES Excellence Cluster).

\appendix
\section*{Appendix}
\section{Conformal mode}\label{conformal mode}

In this appendix, we work in $n=d+1$ spacetime dimensions.
We review how the conformal mode acquires the wrong-sign kinetic term in pure Einstein gravity in the metric formalism.

We start with a Jordan-frame action
\al{
S
	&=	\int\df^nx\sqrt{-g}{M_\P^2\ov2}\Phi R,
		\label{appendix action}
}
where $\Phi =1$ corresponds to pure Einstein gravity.
By a redefinition of the metric
\al{
g_{\mu\nu}
	&=	e^{2\omega}\oc g_{\mu\nu},
}
we obtain
\al{
R
	&=	e^{-2\omega}\pn{\oc R-2\pn{n-1}\oc\Box\omega-\pn{n-2}\pn{n-1}\oc g^{\mu\nu}\p_\mu\omega\p_\nu\omega},&
\sqrt{-g}
	&=	e^{4\omega}\sqrt{-\oc g},
}
where $\oc\Box:=\oc g^{\mu\nu}\oc\nabla_\mu\oc\nabla_\nu$.
The action~\eqref{appendix action}
becomes
\al{
S
	&=	\int\df^nx\sqrt{-\oc g}{M_\P^2\ov2}e^{2\omega}\Phi \pn{
			\oc R-2\pn{n-1}\oc\Box\omega-\pn{n-2}\pn{n-1}\oc g^{\mu\nu}\p_\mu\omega\p_\nu\omega
			}.
}
In the Higgs-Starobinsky inflation, we take $e^{2\omega}\Phi =1$ to switch to the Einstein frame, and the $\oc\Box\omega$ term becomes an irrelevant surface term. Consequently, the kinetic term for $\omega$ has the correct sign.

On the other hand in pure Einstein gravity $\Phi =1$, we have nothing to counter by the conformal factor, and the $\oc\Box\omega$ term yields an extra contribution to the kinetic term when partially-integrated:
\al{
S	&=	\int\df^nx\sqrt{-\oc g}{M_\P^2\ov2}e^{2\omega}\pn{
			\oc R-\pn{n-6}\pn{n-1}\oc g^{\mu\nu}\p_\mu\omega\p_\nu\omega
			}.
			\label{Weyl transformed action}
}
Clearly, we see that $\omega$ has the wrong-sign kinetic term when the number of spacetime dimensions is $1<n<6$.

More precisely, starting from the pure Einstein gravity with $\Phi=1$, written already in the Einstein frame, we may decompose the metric $g_{\mu\nu}$ into the conformal mode $\omega$ and the unimodular part $\widehat g_{\mu\nu}$ within the same Einstein frame~\cite{tHooft:2011aa}:
\al{
g_{\mu\nu}
	&=	e^{2\omega}\widehat g_{\mu\nu},
}
where $\det_{\mu,\nu}\widehat g_{\mu\nu}=-1$.
From Eq.~\eqref{Weyl transformed action}, we see that the conformal mode does have the wrong-sign kinetic term:
\al{
S	&=	\int\df^nx{M_\P^2\ov2}e^{2\omega}\pn{
			\widehat R-\pn{n-6}\pn{n-1}\oc g^{\mu\nu}\p_\mu\omega\p_\nu\omega
			}.
}

\section{Role of Mass Term} \label{App:mass}

\begin{figure} \centering
	\includegraphics[width=0.5\textwidth]{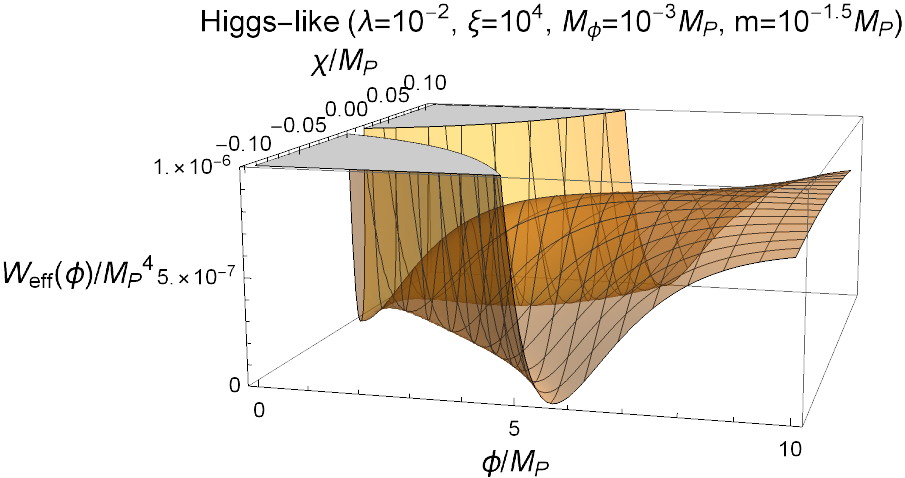}
	\includegraphics[width=0.4\textwidth]{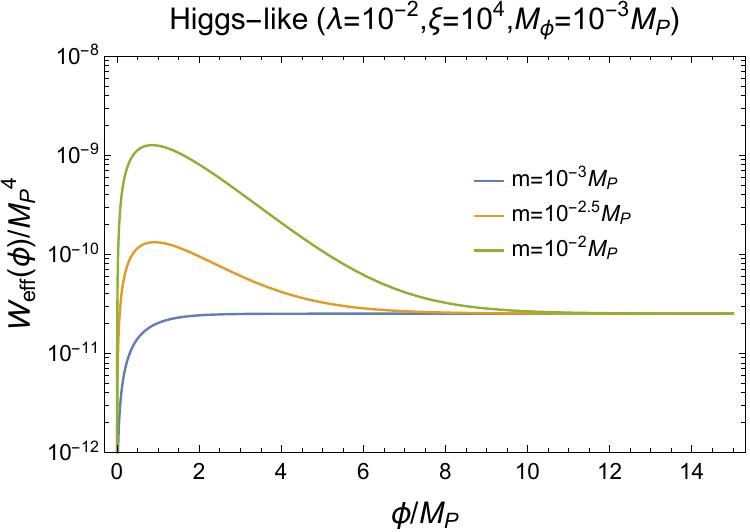}
	\caption{ Figures for the shape of the full potential (left) and the effective potential (right) in the presence of a large mass term with a Higgs-like parameter.  \label{fig:MassPlot} }
\end{figure}

In the presence of the mass term, it turns out that the shape of the potential is deformed even without considering higher dimensional operators.

Following the same procedure obtaining the valley equation with the potential
\begin{align}
	U(\chi) = \frac{1}{2} m^{2} \chi^{2} + \frac{\lambda}{4} \chi^{4},
\end{align}
we have
\begin{align}
	\chi_{\rm v}(\phi) = \begin{dcases}
 \pm M_\xi\sqrt{\frac{ 3{ M_\phi^{2}M_\P^2 } \left(	e^{\sqrt{2\ov3}{\phi\ov M_\P}} - 1	\right) -{m^2\ov M_\xi^2}
	}{\lambda M_\xi^4 + 3{M_\phi^2M_\P^2}}} & \left(  \phi \geq \phi_{\rm crit}  \right) \\
		0 & \left(  \phi < \phi_{\rm crit}  \right)
	\end{dcases}
\end{align}
with
\begin{align}
	\phi_{\rm crit} \equiv \sqrt{\frac{3}{2}} M_{P} \log \left(	1 + \frac{m^{2} M_{\xi}^{4}}{3 M_{\phi}^{2}  M_{\rm P}^{4} }	\right).
\end{align}
Note that, with $ m \neq 0 $, there exists only one valley for small values of $\phi$.

With this expression, effective single field potential becomes
\begin{align}
	&W_{\rm eff} (\phi) \\
	&=
	\begin{dcases}
		\frac{M_\P^4e^{- 2 \sqrt{\frac{2}{3}}{\phi\ov M_\P}}}{4\pn{\lambda+ 3 {M_\phi^2M_\P^2\ov M_\xi^4}}}
	\left[
	3\lambda{M_\phi^{2}\ov M_\P^2}\left( e^{\sqrt{2\ov3}{\phi\ov M_\P} }	- 1\right)^{2}
	+ 6{m^{2}M_\phi^2\ov M_\xi^2M_\P^2} \left( e^{\sqrt{2\ov3}{\phi\ov M_\P} }	- 1\right)
	-{m^{4}\ov M_\P^4}
	\right]  & \left(  \phi \geq \phi_{\rm crit} \right) \\
		\frac{3}{4} M_{\rm P}^{2}M_{\phi}^{2} \left(  1 - e^{- \sqrt{\frac{2}{3}} \phi}	\right)^{2} & \left(  \phi < \phi_{\rm crit}  \right)
	\end{dcases} 
\end{align}
Therefore, for sufficiently large $m$ so that $ \phi_{\rm crit} $ become also large to cover the field range which is relevant for inflation, this model reduces to pure Starobinsky inflation with the scalar $\chi$ decoupled. Taking $\phi_{*} \sim 5 M_{\rm P}$ (corresponding to CMB pivot scale) and $M_{\phi} \sim 10^{-5}M_{\rm P}$ (to fit $A_{s}$), this requires $m > \mathcal{O}(10^{-4}) M_{\rm P}/M_{\xi} $. On the other hand, also for very small $m$, $ \phi_{\rm crit} \simeq 0 $ so it is irrelevant for cosmological observables.

However, for some intermediate value of $m$, especially for Higgs-like regimes, there is a regime where the correction coming from the mass term induces a bump with a local maximum located at
\begin{align}
	\phi_{\rm bump} \equiv \sqrt{\frac{3}{2}} M_{\rm P}	\log \left( 1 + \frac{m^{2} (3M_{\phi}^{2} + m^{2} M_{\xi}^{2}/M_{\rm P}^{2})}{3M_{\phi}^{2}(m^{2} - \lambda M_{\xi}^{2})}	\right).
\end{align}
as depicted in Figure~\ref{fig:MassPlot}. Therefore, if the condition $ \phi_{\rm crit} < \phi_{\rm bump} < \phi_{*} $ is satisfied, this bump radically changes the dynamics of the inflaton and plausibly cosmological observables, while the detailed analysis for whole parameter space is out of the scope of the paper. For Higgs-like parameter, this condition is fulfilled as $ m > \mathcal{O} \left( \sqrt{\lambda} M_{\xi} \right) $. For illustration, we depict the figures for full multi-field potential and the effective single field potential in a Higgs-like parameter varying the parameter $m$ in Figure~\ref{fig:MassPlot}.

In the main text, we basically assumed the mass term is sufficiently small so that these kinds of ambiguity does not arise.

\section{Higgs-$ R^{3} $ inflation \label{app:R3}}
As a straightforward generalization, we consider the cases with different powers of each term. For a specific example, we will consider Higgs+$ R + R^{3} $ Lagrangian in the absence of $ R^{2} $ term. In this Appendix, we set $M_{\rm P} \equiv 1$ for simplicity.

As a more general case, we first consider following $ f\fn{R,\chi} $ and $ U(\chi) $ as
\begin{align}
	f\fn{R,\chi} = \left(1 + \xi \chi^{m}\right)R + \beta R^{n}, && U(\chi) = \frac{\lambda}{4} \chi^{\ell},
\end{align}
giving the effective potential
\begin{align}
	W = e^{- 2 \sqrt{2\ov3}{\phi\ov M_\P}} \left[ \frac{1}{2}  
	\frac{\beta(n-1)}{(n \beta)^{\frac{n}{n-1}}}  \left(e^{\sqrt{2\ov3}{\phi\ov M_\P}} - 1 - \xi \chi^{m} \right)^{\frac{n}{n-1}} + \frac{\lambda}{4} \chi^{\ell} \right].
\end{align}
Hence, one easily notices that $ R^{2} $ case is rather special. Regardless of the form of the non-minimal coupling, $ R^{2} $ potential gives flat potential (as expected because solely $ R^{2} $ term support Starobinsky inflation). On the other hand, higher order $ R^{n} $ with $ n > 2 $ does not have 
flat potential at a large field limit. This just shows the well-known fact that a $ f(R) $-gravity theory with $ R^{n} $ ($ n > 2 $) term is not suitable to realize the potential appropriate for successful inflation. However, the situation changes when there is a large non-minimal coupling between Higgs and Ricci scalar. 
Here we focus on the case of $ n = 3 $, $ m = 2 $, $ l = 4 $ within dimension-4 terms, and will analyze how large the non-minimal coupling should be, and how suppressed the higher dimensional operators should be.

Explicitly, the potential becomes
\begin{align}
	W(\phi,\chi) = \frac{1}{\sqrt{3 \beta}} e^{-2 \sqrt{2\ov3}{\phi}} \chi \left( \sqrt{3 \beta} \lambda \chi^{2} - \xi \left( e^{\sqrt{\frac{2}{3}}{\phi\ov M_\P}} - 1 - \xi \chi^{2} \right)^{1/2} \right).
\end{align}
While its single field reduction form is not particularly illuminating, this has simple expansion in small $ \beta $ limit, as
\begin{align}
	W_{\rm eff} (\phi) \simeq \frac{\lambda}{4 \xi^{2}} \left( 1 - e^{- \sqrt{2\ov3}{\phi}} \right) + \beta \frac{\lambda^{3}}{2 \xi^{6}} e^{\sqrt{\frac{2}{3}}{\phi}} \left(	e^{\sqrt{\frac{2}{3}}{\phi}} - 1	\right)^{3} + \mathcal{O}(\beta^{2})
\end{align}
For the second term proportional to $ \beta $ to be suppressed at the pivot scale, we have to have
\begin{align}
	\frac{\text{(second term)}}{\text{(first term)}} = 2 \beta \frac{\lambda^{2}}{\xi^{4}}	\left(	e^{\sqrt{\frac{2}{3}}{\phi}} -1	\right) \simeq \frac{8}{3} \frac{N_{e} \beta \lambda^{2}}{\xi^{4}}  \ll 1
\end{align}
where $ \phi_{*} $ is the field value correspond the pivot scale with $ N_{e} = 60 $, and we used $ N_{e} \simeq \frac{3}{4} e^{\sqrt{\frac{2}{3}}{\phi}_{*}} \gg 1 $. Hence, we obtain $ \beta \ll \frac{1}{N_{e}}\left(\frac{\xi^{2}}{\lambda}\right)^{2}  $. Similar procedure is easily applied to general $ R^{n} $ case, and we found that $ \beta \ll \frac{1}{N_{e}} \left(	\frac{\xi^{2}}{\lambda} \right)^{n-1} $. We note that the value of $ \xi / \lambda^{2} $ is nearly fixed from the observation of the scalar amplitude.

\begin{figure} \centering
	\includegraphics[width=0.5\textwidth]{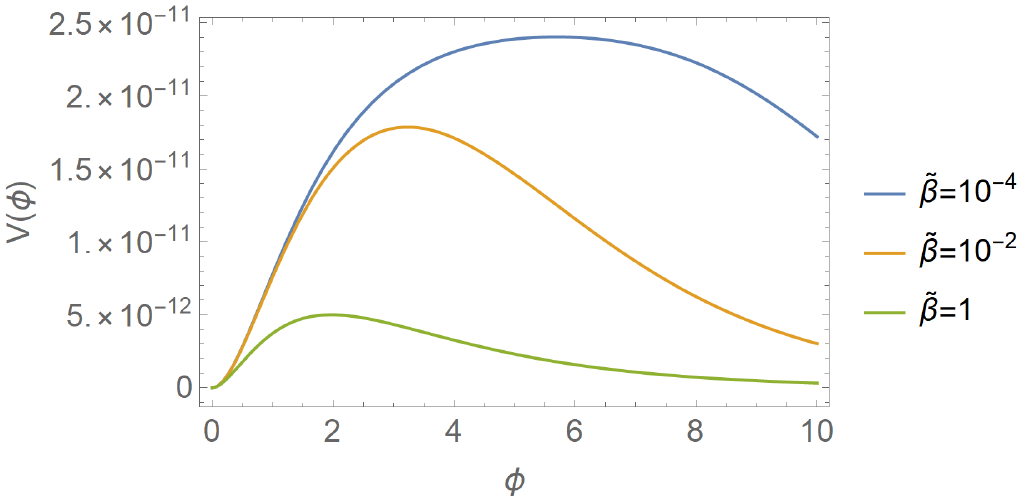}
	\includegraphics[width=0.4\textwidth]{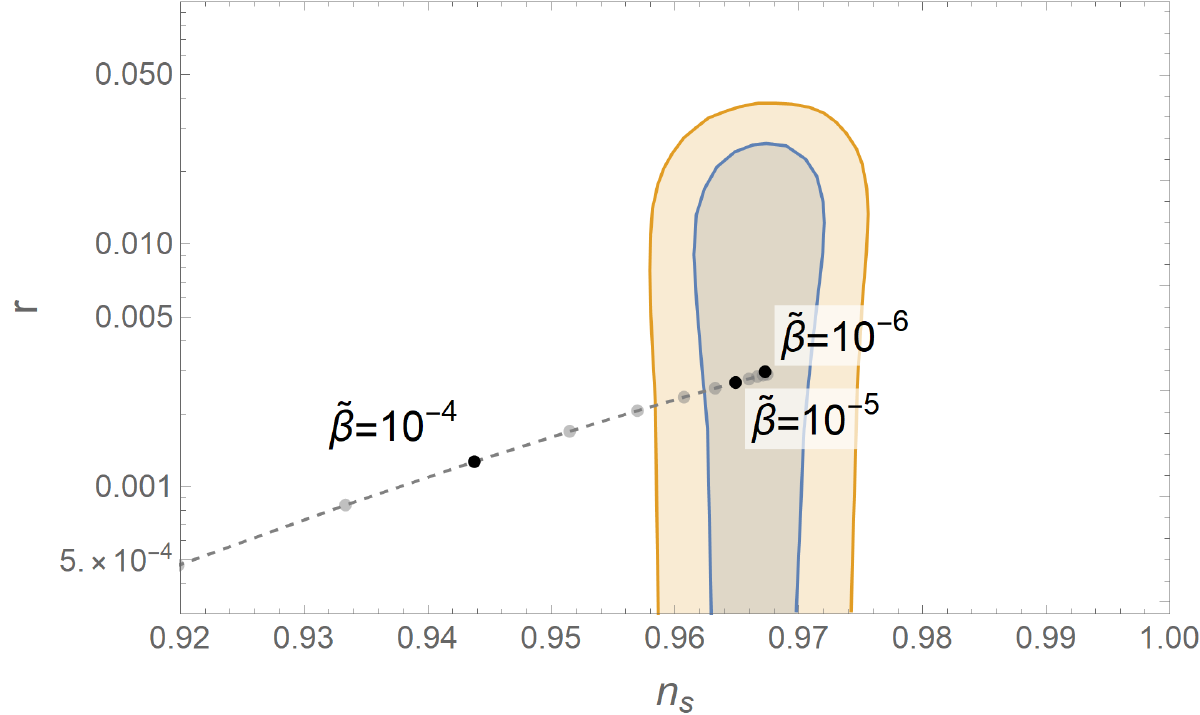}
	\caption{
		(Left) Change of the potential shape for the Higgs-$ R^{3} $ model for various $ \tilde{\beta} $. (Right) Deformation of cosmological observables $ n_{s} $ and $ r $ values in the presence of $R^{3}$ term without $R^{2}$ term.
		\label{figure:R3woR2}}
\end{figure}

To be more precise, we re-parameterize $ \beta = \widetilde{\beta}\left(	\frac{\xi^{2}}{\lambda}	\right)^{2} $ so that $ n_{s} $ and $ r $ only depend on $ \widetilde{\beta} $. Figure~\ref{figure:R3woR2} shows the dependence of the $ n_{s} $ and $ r $ plot with varying $ \tilde{\beta} $. Numerically, we find that $n_s $ and  $r$ are  more sensitive than the naive estimation given above, and it turns out that  
we need 
$ \tilde{\beta} \lesssim 4 \times 10^{-4} $ to guarantee the compatibility to the observation.

\bibliographystyle{utphys}
\bibliography{ftheory.bib}

\end{document}